\documentclass[fleqn,usenatbib]{mnras}

\usepackage{newtxtext,newtxmath}
\usepackage{ulem}
\usepackage{physics}
\usepackage{xcolor}

\usepackage[T1]{fontenc}

\DeclareRobustCommand{\VAN}[3]{#2}
\let\VANthebibliography\thebibliography
\def\thebibliography{\DeclareRobustCommand{\VAN}[3]{##3}\VANthebibliography}

\usepackage{graphicx}	
\usepackage{amsmath}	
\usepackage{ulem}

\title[Width and Magnetization in Early Afterglows]{Ejecta Width and Magnetization Reflected in Gamma-Ray Burst Early Afterglows: Implication for Reverse Shock Component and Shallow Decay Phase}

\author[Y. Kusafuka et al.]{
Yo Kusafuka,$^{1}$\thanks{E-mail: kusafuka@icrr.u-tokyo.ac.jp}
Katsuaki Asano$^{1}$
\\
$^{1}$Institute for Cosmic Ray Research, The University of Tokyo, 5-1-5 Kashiwanoha, Kashiwa, Chiba 277-8582, Japan
}

\date{Accepted XXX. Received YYY; in original form ZZZ}

\pubyear{2024}

\begin{document}
\label{firstpage}
\pagerange{\pageref{firstpage}--\pageref{lastpage}}
\maketitle

\begin{abstract}
To study the ejecta property dependence of the gamma-ray burst (GRB) afterglow, we carry out spherically symmetrical one-dimensional special relativistic magneto-hydrodynamic simulations of magnetized outflows with an adaptive mesh refinement method. 
The Lorentz factor evolutions of forward and reverse shocks induced by the interaction between magnetized ejecta and an ambient medium are investigated for a wide range of magnetization and width of the ejecta. The forward shock evolution is described by the magnetic acceleration, coasting, transition, and self-similar deceleration phases. According to our simulation results, we numerically calculate the corresponding radiation. 
Based on our numerical results, to model afterglow light curves in general cases, we construct semi-analytical formulae for the Lorentz factor evolutions. 
The magnetization and ejecta width dependence are clearly seen in the reverse shock light curves. The transition phase with a reasonable ejecta width can reproduce the shallow decay phase in the observed GRB afterglow. The inverse Compton emission in the magnetic acceleration phase can be responsible for the very steep rise of the early TeV
emission in GRB221009A.
\end{abstract}




\section{Introduction}

Gamma-ray bursts (GRBs) are the most luminous explosions in the Universe. Their enormous energy is promptly released within a few {seconds to} minutes, followed by a gradually fading afterglow radiation.
The energy source of GRBs is thought to be the relativistic jets from compact remnants of the hypernovae or binary neutron star mergers.
Dissipation of the energy through collision {of the jet} with interstellar mediums (ISM) or stellar winds causes multi-wavelength radiation. This external shock model can explain GRB afterglows \citep{1992MNRAS.258P..41R,1995ApJ...455L.143S,1997ApJ...476..232M}.
While the analytical formulae by \citet{1998ApJ...497L..17S} have been frequently adopted for the afterglow lightcurves, several numerical treatments obtained with hydrodynamical simulations \citep[e.g.][]{2012ApJ...749...44V,2020ApJ...896..166R,2020MNRAS.497.4672A,2022MNRAS.510.1315A} or the evolution of the electron energy distribution \citep[e.g.][]{2017ApJ...844...92F,2020ApJ...905..105A} have been also used to constrain the macroscopic and microscopic parameters. However, the {$\sim$40\% \citep{2009MNRAS.397.1177E,2018ApJS..236...26L} of} early X-ray {afterglow} lightcurves show shallower decay \citep{2006ApJ...642..389N,2007ApJ...670..565L} than {what} the standard model predicts. While the energy injection model \citep{2006ApJ...642..354Z} has been considered for the shallow decay phase {observed in the early afterglows}, {the chromatic break of optical lightcurves} is hard to be explained \citep[e.g.][]{2006MNRAS.369..197F,2006MNRAS.369.2059P,2009ApJ...702..489R}. This may require another emission component such as emission from the reverse shock propagating the ejecta \citep[e.g.][]{2014ApJ...781...37P,2014ApJ...785...84J,2019ApJ...879L..26F,2021MNRAS.505.4086G}.

Launching mechanisms of relativistic jets are still debated, but extracting the rotation energy of a compact object by a global magnetic field may be the most promising process \citep{1977MNRAS.179..433B}. 
The magnetic launching mechanism generally produces magnetically dominated jets. Some simulations confirm that magnetically dominated outflows of $\sigma\gg1$ can be launched through the magnetic process {\citep{2019ApJS..243...26P,2022ApJ...933L...2G}}, where $\sigma$ is the magnetization parameter defined as the ratio of the Poynting flux to the enthalpy flux. 
Observationally, optical flashes probably originated from the reverse shock suggest that the jet is mildly or strongly magnetized \citep{2002ChJAA...2..449F,2006ApJ...636L..69W,2006ApJ...652.1416R,2008Natur.455..183R,2013ApJ...774..114J,2015ApJ...798....3Z,2016ApJ...833..100H}.
{Polarization measurements of prompt emission also favor magnetized origin of jets \citep{2022MNRAS.511.1694G,2022ApJ...936...12C,2024arXiv240813199V}, but a spectro-polarimetric analysis of GRB 160325A and GRB 160802A suggests a matter-dominated jet with marginal magnetization \citep{2024ApJ...972..166G}.}
High polarization degrees of $\sim$10\% found in some GRB {afterglows} indicate the existence of the large scale ordered magnetic filed \citep{2003ApJ...583L..63B,2009Natur.462..767S}. 
For the brightest-of-all-time (BOAT) \citep{2023Sci...380.1390L}, ultra-long duration \citep{2024ApJ...962..115R} GRB 221009A, \citet{2023ApJ...947L..11Y,2024JHEAp..41...42Z} claim the magnetic dominance at the dissipation radius. 

The energy dissipation of jets can also occur inside themselves due to their variability. This internal shock model is often used to explain the prompt emission of GRBs and blazar flares \citep{1978MNRAS.184P..61R,1994ApJ...430L..93R,1997ApJ...490...92K,1998MNRAS.296..275D}.
Motivated by the {Imaging X-ray Polarimetry Explorer} \citep{2022JATIS...8b6002W} {observations} for Mrk 501 \citep{2022ApJ...938L...7D} and Mrk 421 \citep{2022Natur.611..677L}, our previous study, \citet{2023MNRAS.526..512K} demonstrates the efficient magnetic energy dissipation by internal shock process.
Magnetic pressure expansion converts the magnetic energy into the kinetic energy of the intermittent ejecta, which is then dissipated into the thermal energy by the shock waves. The dissipation efficiency is almost 10\%, which is comparable to the reported radiative efficiency of GRBs and Blazars \citep{2012Sci...338.1445N}.
However, this in turn suggests that the rest of the $\sim90$\% magnetic energy has remained in the region of the ejecta with $\sigma\gg1$. Thus, magnetically dominated ejecta in the afterglow phase are naturally expected. 

Even after the prompt emission, the ejecta can be accelerated by their remaining magnetic energy \citep{2023MNRAS.526..512K}. This late-phase magnetic acceleration is the same mechanism as the impulsive acceleration \citep{2011MNRAS.411.1323G,2012MNRAS.422..326K}. 
Some numerical simulations show the transition phase from the coasting phase to the self-similar deceleration phase so-called the Blandford-McKee solution \citep{1976PhFl...19.1130B}.
This intermediate phase is due to the initial finite radial width of the ejecta \citep{2000ApJ...542..819K,2009A&A...494..879M,2023MNRAS.526..512K}. {This gradual evolution affects the temporal behaviour of the emission}, but the analytical treatment for that phase is difficult \citep{2013NewAR..57..141G}.
The dynamics of the forward and reverse shocks in this phase may be also affected by the magnetization, which has not been well studied especially for $\sigma\gg1$.

In this paper, we study the effects of the initial magnetization and the radial width of the ejecta on the dynamics of the forward and reverse shocks by performing one-dimensional (1D) special relativistic magneto-hydrodynamical (SRMHD) simulations.
We especially focus on the behaviour of the early afterglow emission including the reverse shock component and shallow decay phase.
The paper is organised as follows. In Section \ref{sec:simulation}, we describe the numerical method for the simulation of relativistic outflows and afterglow radiation. In Section \ref{sec:FS}, we show the results of forward shock radiation by interacting between magnetized ejecta and an ambient medium. In Section \ref{sec:RS}, we show the results of reverse shock radiation. In Section \ref{sec:model}, we provide a semi-analytic model of our simulation results. In Section \ref{sec:afterglow}, we discuss the observational implications of our results for GRB early afterglow. The conclusion is summarised in Section \ref{sec:summary}. 

Throughout the paper, we adopt the flat $\Lambda$CDM cosmology with the parameters from \citet{2020A&A...641A...6P}, where $H_0 = 67.66$ km/s/Mpc, $\Omega_{\rm b}=0.0490$, $\Omega_{\rm m}=0.3111$, and $\Omega_\Lambda=0.6889$. We assume the source is located at $z=1.0$ or $d_{\rm L}=2.1\times10^{28}$ cm, but neglecting the extra-galactic background light (EBL) absorption for simplicity. 
{In most cases, the very high-energy gamma-ray community provides de-absorbed spectra adopting an EBL model in their observation papers.}

\section{Simulation Method}\label{sec:simulation}

\subsection{Fluid Dynamics}

In this paper, we consider magnetically-dominated relativistic ejecta (magnetic bullets) launched from a central engine. To investigate the afterglow radiation from magnetic bullets via interaction with an ambient medium, we carry out 1D SRMHD simulations assuming spherically symmetric geometry. {1D simulations are free from fluid instabilities. In a realistic system, the Rayliegh–Taylor instability may affect on a contact discontinuity to mix the forward and reverse shocked regions. The Kelvin–Helmholtz instability may also contribute to mixing an ejecta with a surrounding medium. For magnetized ejecta, the Kink instability can induce magnetic reconnection to reduce magnetization. We neglect the effects of such instabilities in this paper. We also neglect a lateral structure and a lateral spreading effect of the ejecta, which can affect especially late phase afterglows \citep{1999ApJ...525..737R,2012ApJ...751..155V,2018ApJ...865...94D,2020ApJ...896..166R}}.

Our simulation method is updated from the previous version in \citet{2023MNRAS.526..512K}. {We improve the spatial and time interpolation of variables to achieve a much higher resolution than that in our previous simulations, whose scheme has the 2nd-order interpolation. In addition, we adopt a moving window technique \citep{2004A&A...418..947M} to reduce the computational cost drastically. The details are given in Section \ref{subsubsec:sim_tech}.}

\subsubsection{Fundamental equations}

 We only consider the radial component for the fluid velocity $\vb*{v}=(v,0,0)$ and the $\theta$-component for the magnetic field $\vb*{B}=(0,B,0)$, which are measured at the rest frame of the ambient medium. On the other hand, the mass density $\rho$, gas pressure $p$, and energy density $\epsilon$ are measured at the fluid rest frame. The equation of state is
\begin{equation}
\epsilon=\frac{p}{\hat{\gamma}-1}+\rho c^2,
\label{eq:EoS}
\end{equation}
where $\hat{\gamma}$ is the adiabatic index. 
{As suggested by \citet{2005ApJS..160..199M} taking the equal sign in Taub's inequality \citep{1948PhRv...74..328T}, the equation of state is approximately expressed as $p=(\epsilon^2-\rho^2c^4)/(3\epsilon)$ with correct cold and relativistic asymptotic values \citep{2007MNRAS.378.1118M}. Combining with Eq. (\ref{eq:EoS}), we can derive the adiabatic index as}  
\begin{equation}
\hat{\gamma}=1+\frac{\epsilon+\rho c^2}{3\epsilon}.
\label{eq:index}
\end{equation}
The mass, energy, momentum conservation laws, and the induction equation are described as \citep[e.g.][]{2009A&A...494..879M}
\begin{equation}
\frac{1}{c}\frac{\partial \rho\Gamma}{\partial t}+\frac{1}{r^2}\frac{\partial}{\partial r}\left(r^2\rho \Gamma \beta\right)=0,
\end{equation}
\begin{equation}
\begin{split}
\frac{1}{c}\frac{\partial}{\partial t}&\left[\left(\epsilon+p+\frac{B^2}{4\pi\Gamma^2}\right)\Gamma^2-p-\frac{B^2}{8\pi\Gamma^2}\right]\\
&+\frac{1}{r^2}\frac{\partial}{\partial r}\left(r^2 \left[\left(\epsilon+p+\frac{B^2}{4\pi\Gamma^2}\right)\Gamma^2\beta \right]\right)=0,
\end{split}
\end{equation}
\begin{equation}
\begin{split}
\frac{1}{c}\frac{\partial}{\partial t}&\left[\left(\epsilon+p+\frac{B^2}{4\pi\Gamma^2}\right)\Gamma^2\beta\right]\\
&+\frac{1}{r^2}\frac{\partial}{\partial r}\left(r^2\left[\left(\epsilon+p+\frac{B^2}{4\pi\Gamma^2}\right)\Gamma^2\beta^2+p+\frac{B^2}{8\pi\Gamma^2} \right] \right)=\frac{2p}{r},
\end{split}
\end{equation}
\begin{equation}
\frac{1}{c}\frac{\partial B}{\partial t}+\frac{1}{r}\frac{\partial }{\partial r}\left(r\beta B \right)=0,
\end{equation}
respectively, where $\beta=v/c$ is the fluid velocity normalised by the speed of light, and $\Gamma=1/\sqrt{1-\beta^2}$ is the Lorentz factor of the fluid. For the parameterisation in our simulations, we define the magnetization parameter $\sigma$ as
\begin{equation}
    \sigma\equiv\frac{B^2}{4\pi(\epsilon+p)\Gamma^2}. 
\end{equation}

\subsubsection{Numerical technique} \label{subsubsec:sim_tech}

For the spatial interpolation of variables, we adopt the 7th-order MP7 scheme \citep{1997JCoPh.136...83S}. For the time interpolation of variables, we use the 3rd-order Strong Stability Preserved Runge-Kutta \citep{1998MaCom..67...73G} and set the CFL number around 0.3. We use the minmod function as a flux limiter \citep{1986AnRFM..18..337R}, and compute numerical flux by approximate Riemann solver, the CENTRAL scheme \citep{KURGANOV2000241}, or so-called Rusanov scheme \citep{RUSANOV1962304}. For the primitive recovery, we use the Newton-Rhapson method \citep{2006MNRAS.368.1040M}. We implement the adaptive mesh refinement \citep[AMR]{1984JCoPh..53..484B} to obtain a higher resolution around discontinuities. To reduce enormous computational costs, we use a moving window technique \citep{2004A&A...418..947M} to follow only around shock waves in the ejecta. To minimise artificial boundary effects, we set the box size to 20 times larger than the initial width of the ejecta. 

{
In high $\sigma$ cases, an insufficient resolution will affect the reverse shock dynamics, making the shock crossing time longer \citep{2023MNRAS.526..512K}. We have confirmed that our resolution is enough to capture the reverse shock dynamics accurately by our means of resolution study. }

\subsubsection{Initial setup}

The simulation window is in the ambient medium rest frame. For the ambient medium, we assume that the magnetic field and the number density are homogeneous and set as $1\ \mu$G and $n_0=1\ \rm{cm}^{-3}$, respectively. {The magnetic field configuration in a typical GRB environment may not be homogeneous, but the field in the ambient medium is so weak that the forward shock dynamics is not affected.
In addition, as usually assumed, we amplify the magnetic field in the forward shocked region adopting the microscopic parameter $\epsilon_B$ to calculate the emission property.} The temperature is also fixed as a non-relativistic temperature of 1 MeV to stabilize the simulations, {but it is sufficiently cold not to affect the evolution of the dynamics}. 

The initial magnetic field of the ejecta is written as
\begin{equation}
    B=\sqrt{4\pi(\epsilon+p)\Gamma^2_0\sigma_0},
\end{equation}
where $\sigma_0$ and $\Gamma_0=10$ are the initial magnetization and the Lorentz factor of the ejecta, respectively. We define the deceleration radius $R_{\rm dec}$ as \citep{1995ApJ...455L.143S}
\begin{equation}
    R_{\rm dec}=\left(\frac{3E_0}{4\pi n_0m_{\rm p}c^2\Gamma_0^2}\right)^{1/3}.
    \label{eq:rdec}
\end{equation}
The inner edge of the ejecta is initially set at $R_0=0.1R_{\rm dec}$. 
We consider two cases for the initial width of the ejecta: thick shell case $\Delta_0=R_{\rm dec}/\Gamma^2_0$ and thin shell case $\Delta_0=R_0/\Gamma^2_0$. 
We assume that the ejecta has initially homogeneous magnetization $\sigma_0$, temperature $T_0=100$ MeV, and Lorentz factor $\Gamma_0$, but the mass density follows as $\rho=\rho_0(r/R_0)^{-2}$ in the initial condition. 
The initial energy of the ejecta is set to be $E_0=10^{50}$ erg, from which
the normalisation of the mass density $\rho_0$ can be calculated.
The deceleration radius $R_{\rm dec}$ is $5.4\times10^{16}$ cm for $E_0=10^{50}$ erg, $\Gamma_0=10$, and $n_0=1\ \rm{cm}^{-3}$.

The simulation window size is 20$\Delta_0$. 
The outer boundary is a moving boundary, where a simulation window moves one cell outer when the forward shock is at the 50 cells before the outer boundary. 
The inner boundary is set to be an injection boundary, where we inject a plasma with a significantly low luminosity with $\Gamma=2$ continuously to avoid a numerical floor. The initial inner boundary is at 500 static cells before the position of the rear edge of the ejecta.
The initial width of the ejecta is resolved by $10^3$ static cells and effectively almost $10^4$ static cells.

\subsection{Radiation}\label{sec:radiation}

Based on our simulation results of the dynamics of shock waves, we calculate radiation from the shocks as a post-process with the one-zone approximation. From the simulation results, we detect the forward and reverse shock radii $R$. Using the physical quantities just behind the shocks, the density $n$, internal energy density $\epsilon$, Lorentz factor $\Gamma$, and magnetic field $B'=B/\Gamma$, we calculate radiation from the two shock fronts. We consider two emission mechanisms: synchrotron radiation and synchrotron self-Compton (SSC) emission. In addition, we take into account two absorption processes: synchrotron self-absorption (SSA) and $\gamma\gamma$ annihilation. Some hadronic emissions such as proton synchrotron and pion production might be dominant in the early phase afterglow, but we leave these effects for future studies. 

\subsubsection{Non-thermal particle spectra}

At or around shock waves, charged particles are accelerated through diffusive shock acceleration (DSA) \citep{1978ApJ...221L..29B,1978MNRAS.182..147B}, magnetic reconnection \citep{2011ApJ...726...75S,2015MNRAS.450..183S}, or turbulent acceleration \citep{2009ApJ...705.1714A,2014ApJ...780...64A}. 
We assume that electrons are injected with a single power law distribution as \citep{1998ApJ...497L..17S}
\begin{equation}
    n_{\rm inj}(\gamma)\propto\gamma^{-p},\ \ \ \ \ \gamma_{\rm m}\leqq\gamma\leqq\gamma_{\rm M},
    \label{eq:Pinj}
\end{equation}
where $n_{\rm inj}(\gamma)$ is the comoving spectral number density at injection as a function of the Lorentz factor $\gamma$ of electrons. 
{Non-relativistic DSA theory expects the spectral index becomes $p=2.0$ in the strong shock limit \citep{1978ApJ...221L..29B,1978MNRAS.182..147B}. However, in the case of relativistic shocks, the index becomes $p\sim2.2$ in the same limit \citep{2005PhRvL..94k1102K}. 
Thus, we set the spectral index $p$ as a fixed value $p=2.2$ for simplicity in this study. In general, the index may be larger than 2.2, making a radiation spectrum softer \citep{1998ApJ...497L..17S}.}
The normalization of $n_{\rm inj}(\gamma)$ and the minimum Lorentz factor $\gamma_{\rm m}$ are determined by assuming that all electrons behind the shock front are accelerated, and obtain an energy fraction $\epsilon_{\rm e}$ ($=0.1$ in our simulations) to the internal energy density $\epsilon$ 
\citep{1998ApJ...497L..17S}: 
\begin{equation}
    \gamma_{\rm m}=1+\frac{p-2}{p-1}\frac{\epsilon_{\rm e} \epsilon}{nm_{\rm e}c^2},
    \label{eq:gamma_m}
\end{equation}
The maximum Lorentz factor is given by the balance of acceleration with the Bohm limit and synchrotron cooling, or the confinement condition \citep{1984ARA&A..22..425H}:
\begin{equation}
    \gamma_{\rm M}=\min\left( \sqrt{\frac{6\pi e}{\sigma_{\rm T}B'}}, \frac{eB' \Delta R}{m_{\rm e}c^2} \right),
    \label{eq:gamma_M}
\end{equation}
where $\sigma_{\rm T}$ is Thomson scattering cross section, $e$ is the elementary charge, $m_{\rm e}$ is the electron mass. 
The shell width of the forward shock $\Delta R \equiv R_{\rm FS}/(12 \Gamma_{\rm FS})$ is estimated from the simulation results of the forward shock radius $R_{\rm FS}$ and the Lorentz factor $\Gamma_{\rm FS}$.
The shell width of the reverse shock $\Delta R\equiv\Gamma_{\rm RS}(R_{\rm CD}-R_{\rm RS})$ is estimated from the simulation results of the contact discontinuity radius $R_{\rm CD}$, the reverse shock radius $R_{\rm RS}$, and the Lorentz factor $\Gamma_{\rm RS}$.
For the forward shock emission, we assume that the magnetic field amplification is due to kinetic instabilities (e.g. Weibel instability). An energy fraction $\epsilon_{\rm B}$ ($=0.01$ in our simulations) to the internal energy density $\epsilon$ can be transferred to the generated turbulent magnetic field energy \citep{1998ApJ...497L..17S}: 
\begin{equation}
    B'=\sqrt{8\pi \epsilon_{\rm B}\epsilon }.
    \label{eq:epsilon_B}
\end{equation}
For the reverse shock emission, the magnetic field is adopted from the simulation date. 

The spectra of these non-thermal particles are modified by radiative cooling. 
The continuity equation of electrons is
\begin{equation}
    \frac{\partial n(\gamma)}{\partial t}+\frac{\partial}{\partial \gamma}\left(n(\gamma)\frac{d\gamma}{dt}\right)=Q,
    \label{eq:CE}
\end{equation}
where $Q$ is the injection term estimated from Eq. (\ref{eq:Pinj}). The radiative cooling function is given by
\begin{equation}
    \frac{d\gamma}{dt}=-\frac{\sigma_T{B'}^2}{6\pi mc}{\gamma}^2\left[1+Y(\gamma)\right],
\end{equation}
where $Y$ is the Compton parameter defined as the ratio of SSC power and synchrotron power.
Due to the Klein-Nishina effect, the Compton $Y$ has energy dependence \citep{2006MNRAS.369..197F}
\begin{equation}
    Y(\gamma)=\frac{-1+\sqrt{1+4\epsilon_{\rm rad}\eta_{\rm KN}\epsilon_{\rm e}/\epsilon_{\rm B}}}{2},
\end{equation}
where $\epsilon_{\rm rad}=\min\left(1,\left(\gamma_{\rm m}/\gamma_{\rm c}\right)^{(p-2)}\right)$ is the fraction of radiated electron energy \citep{2001ApJ...548..787S}, and $\eta_{\rm KN}$ is Klein-Nishina correction factor. 
For fast cooling case ($\gamma_{\rm c}<\gamma_{\rm m}$),
\begin{equation}
    \eta_{\rm KN}=
    \left\{
    \begin{array}{lll}
    0\ \ \ \ \ \ \ \ \ \ \ \ \ \ \ \ \ \ \ \ \ \ \ \ \ \ \ \ \ \ \ \ \ \ \ \ \ \ \ \ \ \ \ \hat{\gamma}<\gamma_{\rm c}\\
    \frac{\hat{\gamma}-\gamma_{\rm c}}{[(p-1)/(p-2)]\gamma_{\rm m}-\gamma_{\rm c}}\ \ \ \ \ \ \ \ \ \ \gamma_{\rm c}<\hat{\gamma}<\gamma_{\rm m}\\
    1-\frac{\gamma_{\rm m}^{p-1}\hat{\gamma}^{2-p}}{(p-1)\gamma_{\rm m}-(p-2)\gamma_{\rm c}}\ \ \ \ \ \ \ \ \gamma_{\rm m}<\hat{\gamma}
    \end{array}
    \right.,
\end{equation}
while for slow cooling case ($\gamma_{\rm m}<\gamma_{\rm c}$), 
\begin{equation}
    \eta_{\rm KN}=
    \left\{
    \begin{array}{lll}
    0\ \ \ \ \ \ \ \ \ \ \ \ \ \ \ \ \ \ \ \ \ \ \ \ \ \ \ \ \ \ \ \ \ \ \ \ \ \ \ \ \hat{\gamma}<\gamma_{\rm m}\\
    \frac{\hat{\gamma}^{3-p}-\gamma_{\rm m}^{3-p}}{[1/(p-2)]\gamma_{\rm c}^{3-p}-\gamma_{\rm m}^{3-p}}\ \ \ \ \ \ \ \ \gamma_{\rm m}<\hat{\gamma}<\gamma_{\rm c}\\
    1-\frac{(3-p)\gamma_{\rm c}\hat{\gamma}^{2-p}}{\gamma_{\rm c}^{3-p}-(p-2)\gamma_{\rm m}^{3-p}}\ \ \ \ \ \ \ \ \gamma_{\rm c}<\hat{\gamma}
    \end{array}
    \right.,
\end{equation}
where  $\hat{\gamma}=\sqrt{4\pi m_{\rm e}c\hat{\nu'}/(3eB')}$ and $\hat{\nu'}$ is given by $\gamma h\hat{\nu'}=\Gamma m_{\rm e}c^2$ \citep{2006MNRAS.369..197F,2024ApJ...962..115R}.

We take the MHD snapshot in every $10^4$ s, which is shorter than the dynamical timescale $R/(\Gamma c)$ at any time. 
This guarantees the particle distribution is steady in each snapshot, and then the corresponding particle spectra are analytically calculated from Eq. (\ref{eq:CE}). 
For the fast cooling case ($\gamma_{\rm c}<\gamma_{\rm m}$) \citep{2009ApJ...703..675N,2010ApJ...712.1232W},
\begin{equation}
    n(\gamma)\propto
    \frac{1}{1+Y(\gamma)}
    \left\{
    \begin{array}{ll}
    {\gamma}^{-2}\ \ \ \ \ \ \gamma_{\rm c}<\gamma<\gamma_{\rm m}\\
    {\gamma}^{-p-1}\ \ \ \ \gamma_{\rm m}<\gamma
    \end{array}
    \right.,
\end{equation}
while for the slow cooling case ($\gamma_{\rm m}<\gamma_{\rm c}$),
\begin{equation}
    n(\gamma)\propto
    \left\{
    \begin{array}{ll}
    {\gamma}^{-p}\ \ \ \ \ \ \ \ \ \ \ \ \ \ \ \ \ \ \ \gamma_{\rm m}<\gamma<\gamma_{\rm c}\\
    \frac{1}{1+Y(\gamma)}{\gamma}^{-p-1}\ \ \ \ \ \ \gamma_{\rm c}<\gamma
    \end{array}
    \right.,
\end{equation}
where the cooling break energy $\gamma_{\rm c}$ is determined by the balance of radiative cooling timescale and dynamical timescale:
\begin{equation}
    \gamma_{\rm c}=\frac{6\pi\Gamma m_{\rm e}c^2}{\sigma_{\rm T}{B'}^2R(1+Y(\gamma_c))}.
    \label{eq:gamma_c}
\end{equation}

\subsubsection{Translation to the Observer Frame}

Due to the relativistic Doppler effect, the observed frequency is boosted as
\begin{equation}
    \nu_{\rm obs}=\mathcal{D} \frac{\nu'}{1+z},
    \label{eq:nu_obs}
\end{equation}
where $z$ is the cosmological redshift and $\mathcal{D}$ is the Doppler factor:
\begin{equation}
    \mathcal{D}(\theta)=\frac{1}{\Gamma(1-\beta\cos\theta)},
    \label{eq:Doppler}
\end{equation}
representing the angle between the line of sight and the direction of the fluid motion by $\theta$. 

We also consider the Equal Arrival Time Surface (EATS) for the observed radiation flux \citep{1997ApJ...491L..19W,2005ApJ...631.1022G}. We define the observed time $t_{\rm obs}$ at which photons emitted from a position at $(R,\theta)$ arrive at the observer,
\begin{equation}
    t_{\rm obs}\equiv(1+z)\left[t-\frac{R(t)-R_0}{c}\cos\theta\right],
    \label{eq:t_obs}
\end{equation}
where $t$ is measured at the engine rest frame. To compute the observed radiation flux, we should integrate over solid angle $d\Omega=2\pi\sin\theta d\theta$. 
The relativistic beaming effect allows us to integrate within a small angle. In our calculation, the maximum angle for the integral is chosen as $\theta_0=3\Gamma^{-1}$.

\subsubsection{Synchrotron radiation}

To calculate the observed synchrotron flux $F_\nu$, we first evaluate the comoving synchrotron emissivity $j'_{{\rm syn},\nu'}$ as \citep{1986rpa..book.....R}
\begin{equation}
    j'_{{\rm syn},\nu'}=\frac{1}{4\pi}\int_{\gamma_{\rm m}}^{\gamma_{\rm M}} d\gamma\  n(\gamma)\frac{\sqrt{3}e^3B'\sin\alpha}{m_{\rm e}c^2} \mathcal{F}\left(\frac{\nu'}{\nu'_0}\right),
\end{equation}
where we use the averaged value of $\sin\alpha=\pi/4$, the typical synchrotron frequency $\nu'_0=3eB'{\gamma_{\rm e}}^2/(4\pi m_{\rm e}c)$, and the fitting formulae for calculating the non-dimensional function $\mathcal{F}(x)$ \citep{2013RAA....13..680F}. 
Then, the isotropic-equivalent flux is written as
\begin{equation}
\begin{split}
    F_{{\rm syn},\nu}(t_{\rm obs})=(1+z) \int \frac{R^2d\Omega}{d_{\rm L}^2} \mathcal{D}^3\Delta R \frac{j'_{{\rm syn},\nu'}}{\tau_{\rm SSA}}\left( 1-\exp(-\tau_{\rm SSA}) \right),
    \label{eq:syn}
\end{split}
\end{equation}
where $d_{\rm L}$ is the luminosity distance, and $R^2 \Delta R$ is a function of $t_{\rm obs}$ and $\theta$ given by the EATS. 
The optical depth due to SSA $\tau_{\rm SSA}=\alpha_\nu \Delta R$ is written with the absorption coefficient
\begin{equation}
    \alpha_\nu=\frac{-1}{8\pi m_{\rm e}\nu^2}\int d\gamma\frac{\sqrt{3}e^3B\sin\alpha}{m_{\rm e}c^2}\mathcal{F}\left(\frac{\nu}{\nu_0}\right)\gamma^2\frac{\partial}{\partial\gamma}\left(\frac{n(\gamma)}{\gamma^2}\right).
\end{equation}

\subsubsection{Synchrotron self-Compton process}

We calculate the comoving SSC emissivity $j'_{{\rm SSC},\nu'}$ as \citep{2008MNRAS.384.1483F}
\begin{equation}
    j'_{{\rm SSC},\nu'}=\frac{1}{4\pi}\int h\nu' n_{\rm ph}(\gamma) n(\gamma)d\gamma,
\end{equation}
where $h$ is the Planck constant, and $n_{\rm ph}(\gamma)$ is the scattered photon spectrum per electron.
\begin{equation}
    n_{\rm ph}(\gamma)=\int d\nu'_{\rm syn} \frac{3\sigma_{\rm T} cn_{\nu'_{\rm syn}}}{4\gamma^2\nu'_{\rm syn}} F_{\rm KN}(g,q).
\end{equation}
The typical frequency of a scattered photon is
\begin{equation}
    \nu'=\frac{\gamma^2\nu'_{\rm syn}}{1+g},
\end{equation}
where $g$ is the threshold for the Klein-Nishina effect defined by
\begin{equation}
    g\equiv\frac{\gamma h\nu'_{\rm syn}}{m_{\rm e}c^2}.
\end{equation}
The comoving synchrotron photon number density is written as
\begin{equation}
    n_{\nu'_{\rm syn}}\simeq\frac{4\pi j'_{\nu'_{\rm syn}}}{h\nu'_{\rm syn}} \frac{\Delta R}{c},
\end{equation}
The function $F_{\rm KN}(g,q)$ is a correction factor of the Klein-Nishina effect \citep{1970RvMP...42..237B}.
\begin{equation}
    F_{\rm KN}(g,q)=2q\ln q+(1+2q)(1-q)+\frac{8g^2q^2}{1+4gq}(1-q),
\end{equation}
where we define
\begin{equation}
    f\equiv\frac{h\nu'}{\gamma m_{\rm e}c^2},
\end{equation}
\begin{equation}
    q\equiv\frac{f}{4g(1-f)}.
\end{equation}

Very high-energy gamma rays are usually absorbed through pair creation. The cross-section for $\gamma\gamma$ annihilation is given by \citep{1967PhRv..155.1404G}
\begin{equation}
    \sigma_{\gamma\gamma}=\frac{3\sigma_{\rm T}}{8s^2}\left[\left(2s+2-s^{-1}\right)\ln(\sqrt{s}+\sqrt{s-1}) - (s+1)(1-s^{-1}) \right],
    \label{eq:annihilation}
\end{equation}
where
\begin{equation}
    s\equiv\frac{\epsilon_1\epsilon_2}{2m_{\rm e}^2c^4},
\end{equation}
is the normalised energy interacting between two photons with energy $\epsilon_1$ and $\epsilon_2$. Because we assume the target photon is distributed isotropic for simplicity, the expression of Eq. (\ref{eq:annihilation}) has no information about the scattering angle.
Then, the optical depth of the pair creation can be calculated as
\begin{equation}
    \tau_{\gamma\gamma}(\epsilon_1)=\Delta R\int_{\epsilon(s>1)}^\infty d\epsilon_2\sigma_{\gamma\gamma}(\epsilon_1,\epsilon_2)n_{\rm \nu'_{\rm syn}}(\epsilon_2).
\end{equation}
The lower bound of the energy integration corresponds to the minimum interaction energy threshold.
We neglect radiation from secondary particles for simplicity. Then, the isotropic-equivalent flux is written as
\begin{equation}
\begin{split}
    F_{{\rm SSC},\nu}(t_{\rm obs})=(1+z)\int \frac{R^2d\Omega}{d_{\rm L}^2} \mathcal{D}^3\Delta R\frac{j'_{{\rm SSC},\nu'}}{\tau_{\gamma\gamma}}\left( 1-\exp(-\tau_{\gamma\gamma}) \right).
    \label{eq:SSC}
\end{split}
\end{equation}

\section{Forward Shock Emission} \label{sec:FS}

In this section, we study effects of the magnetic acceleration on the dynamics and radiation from the forward shock. 
We simulate a single relativistic ejecta plunging into a homogeneous low-magnetized ambient medium. 
The interaction between the ejecta and the medium generates a strong forward shock propagating into the medium. 
The initial magnetic energy of the ejecta is converted into the kinetic and thermal energy of the shocked ambient medium. 
This energy transfer is due to the magnetic pressure expansion \citep{2023MNRAS.526..512K}. 

{Simulations with a very high $\Gamma$ (such as $\Gamma=100$) are very hard to carry out. First, we simulate the shocks with $\Gamma_0=10$. Then, based on the simulation results, We attempt to establish a semi-analytic formulation for a higher $\Gamma$ in Section \ref{sec:model}, and discuss observational appearances in Section \ref{sec:afterglow}.}

\subsection{Dynamics of the forward shock\label{subsec:FS_ISM}}

\begin{figure}
\includegraphics[width=\columnwidth]{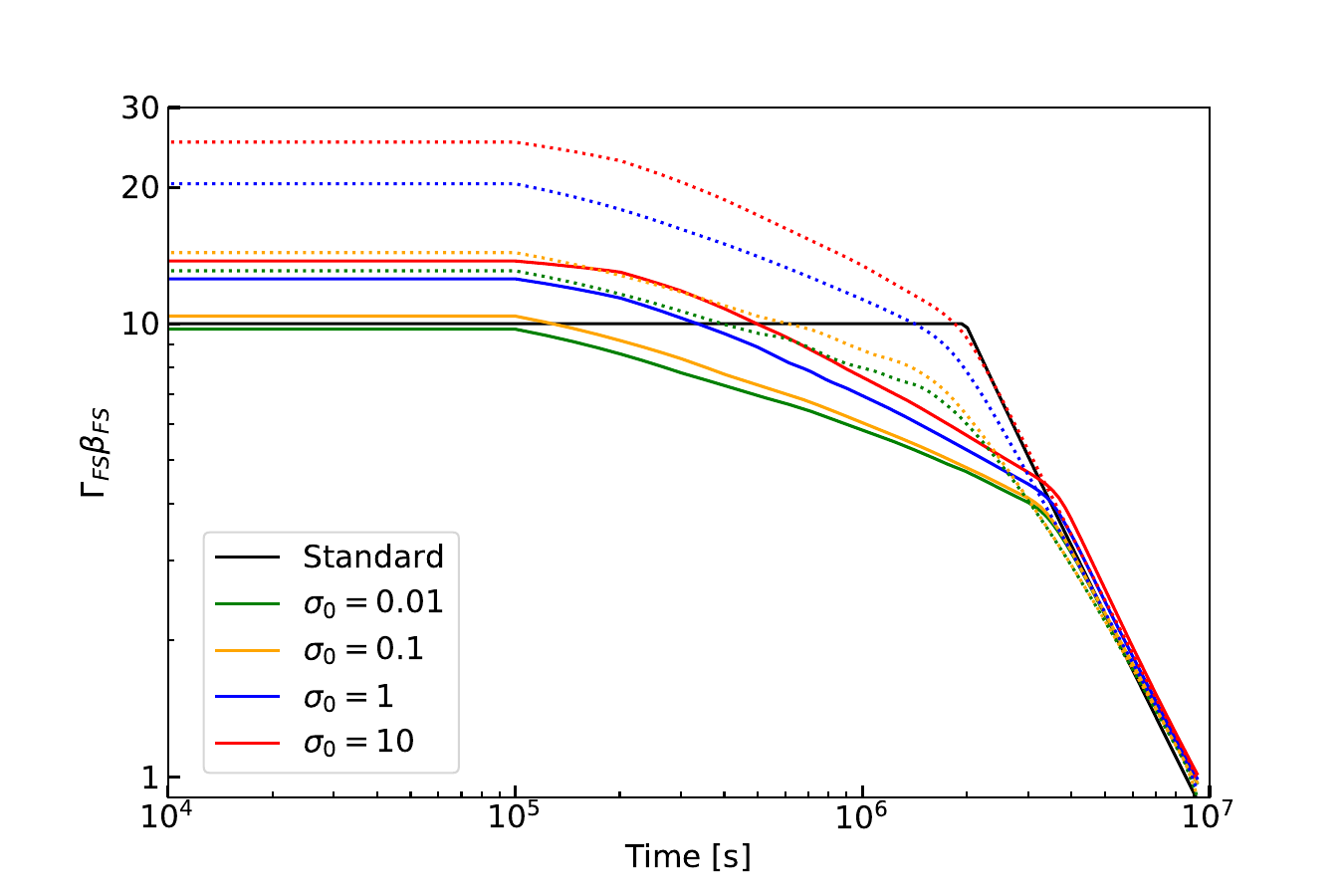}
\caption{The time evolution of $\Gamma\beta$ of the shocked ambient medium just behind the forward shock.
The parameters are $E_0=10^{50}$ erg, $\Gamma_0=10$, and $n_0=1\ \rm{cm}^{-3}$.
The solid line represents the thick shell case and the dotted one represents the thin shell case. Each coloured line corresponds to a different initial magnetization $\sigma_0$. The black line shows the frequently used approximation (standard model) in the thin shell case. The break time scale of the standard model is the deceleration time $t_{\rm dec}$ defined by Eq. (\ref{eq:rdec}). 
\label{fig:FS}}
\end{figure}

\begin{figure}
\includegraphics[width=\columnwidth]{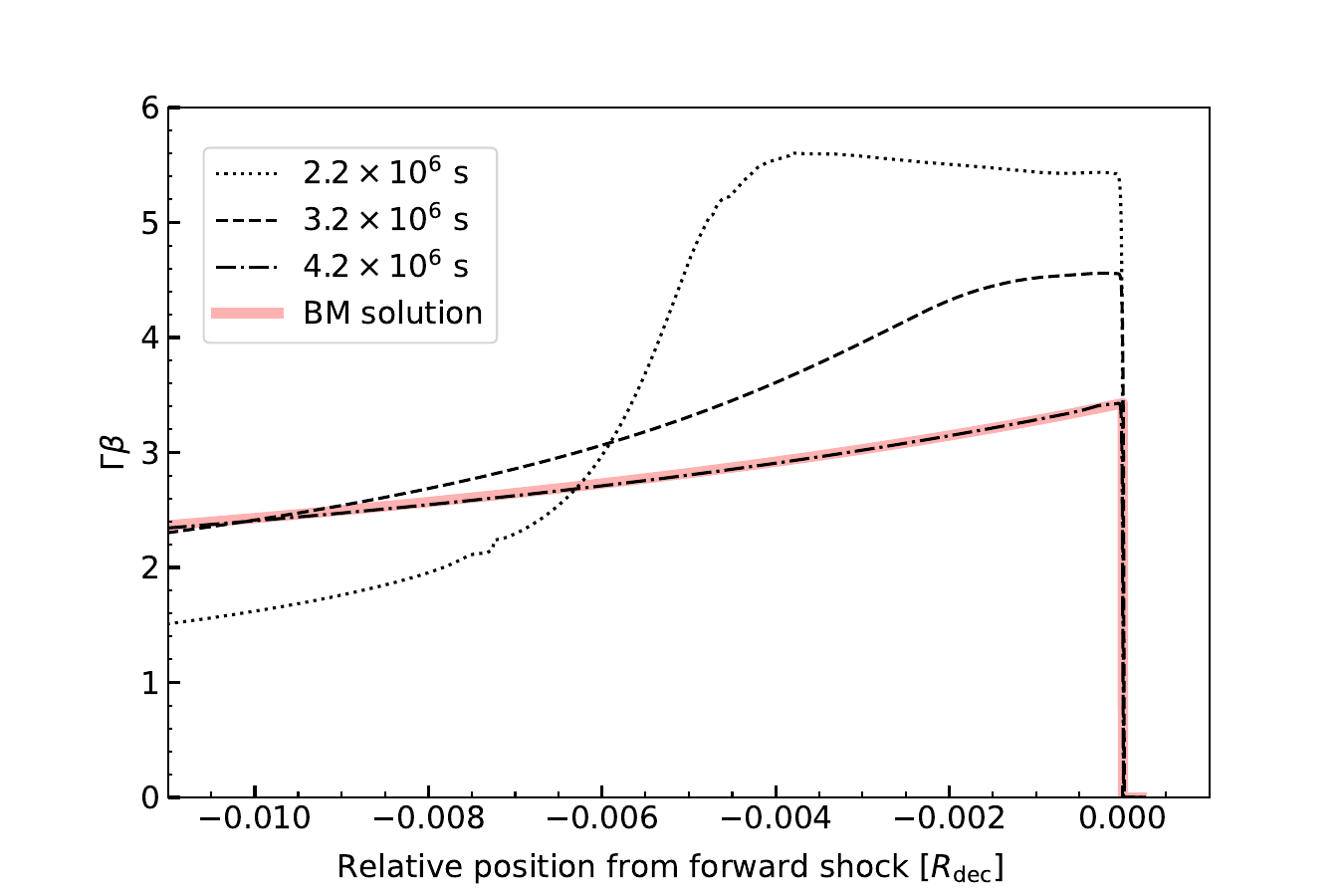}
\caption{The radial profiles of $\Gamma\beta$ around the end of the transition phase for $\sigma_0=10$ {showing the temporal evolutution. 
The rarefaction wave from the rear edge of the ejecta approaches to the forward shock front, finally achieving the BM solution as the red mask expressed.}
\label{fig:RF}}
\end{figure}

In the simplest approximation, until the deceleration time scale $t_{\rm dec}\simeq R_{\rm dec}/c$, the dynamics of the forward shock is free expansion with constant speed. Beyond $t_{\rm dec}$, the forward shock evolves adiabatically with decaying $\Gamma \beta$ given by the Blandford-McKee (BM) solution \citep{1995ApJ...455L.143S,2005ApJ...628..315Z,2008A&A...478..747G}. The BM solution is a self-similar solution for an ultra-relativistic outflow from a point-like explosion. According to this solution, the Lorentz factor of the shock front $\Gamma_{\rm FS}$ evolves as $\Gamma_{\rm FS} \propto t^{-3/2}\propto t_{\rm obs}^{-3/8}$ \citep{1976PhFl...19.1130B}.
For the thick shell case \citep{1995ApJ...455L.143S}, the propagation of the reverse shock affects the evolution of $\Gamma_{\rm FS}$ before the BM stage as $\Gamma_{\rm FS} \propto t^{-1/2}\propto t_{\rm obs}^{-1/4}$ \citep{1997ApJ...489L..37S}.
Hereafter, we denote the above description of the Lorentz factor evolution as the "standard" model. 

Our numerical results in Figure \ref{fig:FS} show different time evolutions of $\Gamma_{\rm FS}$ from the standard model, where $\Gamma_{\rm FS}$ is estimated at the radius where the gas pressure becomes maximum behind the shock front. 
The differences in the initial Lorentz factor are due to the impulsive acceleration \citep{2011MNRAS.411.1323G}: the magnetic pressure accelerates the head of the ejecta. 
The acceleration of the forward shock lasts until the reverse shock ignition. A detailed description of the transition from the acceleration phase to the coasting phase will be given in \S \ref{sec:ev-tran}.

Unfortunately, the forward shocked region cannot be resolved in the early stage of the coasting phase. However, the reverse shock is resolved so that the Lorentz factor at the end of the coasting phase $\Gamma_{\rm i}$ is roughly consistent with the analytical estimate in \S \ref{sec:ev-tran}. 
To calculate radiation from the forward shock in the coasting stage, we interpolate physical quantities such as the Lorentz factor. As the shocked region is resolved just after the end of the coasting phase,
the Lorentz factor is interpolated as a constant $\Gamma_{\rm i}$ before this time following the definition of the coasting phase. The number density $n_{\rm FS}$ and internal energy density $\epsilon_{\rm FS}$ in the coasting phase are obtained from the Rankine-Hugoniot relation \citep{1976PhFl...19.1130B} as
\begin{flalign}
    &n_{\rm FS}=(4\Gamma_{\rm i}+3) n_0, \\
    &\epsilon_{\rm FS}=(4\Gamma_{\rm i}+3)(\Gamma_{\rm i}-1)n_0m_{\rm p}c^2.
\end{flalign}

Our results show two breaks for both the thin and thick shell cases.
The first break at the end of the coasting phase appears an order of magnitude faster than the deceleration time (the break time for the standard model).
The coasting phase ends when the density of the reverse shocked region decreases to $\sim \Gamma_{\rm i}^2 n_0/(1+\sigma_{\rm RS})$, where the subscript ${\rm RS}$ denotes the quantity in the reverse shocked region (see discussion in \S \ref{sec:ev-tran}). This transition radius $R_{\rm T}$ is similar to the radius $R_{\rm N}$ defined in \citet{1995ApJ...455L.143S}; the reverse shock becomes relativistic at $R=R_{\rm N}$ in non-magnetized cases. In high-$\sigma$ cases, the reverse shock is already relativistic, and the reverse shock crossing radius $R_\Delta$ can be smaller than $R_{\rm T}$ even in the thick shell case.

As discussed in \citet{2023MNRAS.526..512K}, from $R=R_{\rm T}$ to the second break at $R=R_{\rm BM}$, the deceleration is gradual compared to the BM solution owing to the finite thickness of the ejecta. Figure \ref{fig:RF} shows the profile of $\Gamma\beta$ 
just before the second break.
At $t=2.2 \times 10^6$ s, the reverse shock already crossed the ejecta. The rarefaction wave gradually catches up with the forward shock front. During this phase, the deceleration of the forward shock is suppressed by the pressure of the reverse shocked region.
We call this phase the "transition phase" ($R_{\rm T}<R<R_{\rm BM}$), which is different from the reverse shock crossing phase between $R_{\rm N}$ and $R_\Delta$. This $\Gamma\beta$-evolution in the transition phase will modify light curves in GRB early afterglows.

The dynamics in the transition phase might be described by BM solution with internal energy injection \citep{2006ApJ...642..354Z,2013NewAR..57..141G,2014MNRAS.442.3495V} as
\begin{equation}
    \Gamma_{\rm FS}\propto t^{-\frac{1}{3+s}},
\end{equation}
where $s$ denotes the index for the energy increase with the decelerating $\Gamma$, 
$E\propto\Gamma^{1-s}$. 
According to our results, the slope of the transition phase has a slight dependence on the initial magnetization as $\Gamma_{\rm FS}\propto t^{-\alpha(\sigma_0)}$. 
The index $\alpha(\sigma_0)=3/(1+s)$ ranges from $0.32$ to $0.48$ as $\sigma_0$ increases. 
This in turn means the injection rate $s$ ranges from $5$ to $8$ as $\sigma_0$ decreases. 
This magnetization dependence results in the different rising slopes in the light curve of early afterglows. The energy injection rate can be estimated by the pressure balance at the contact discontinuity. 
We will derive the analytical expression of the transition phase in Section \ref{sec:model}. 

The transition phase lasts until the rarefaction wave from the rear edge of the ejecta reaches the forward shock front at $R=R_{\rm BM}$ \citep{2000ApJ...542..819K,2009A&A...494..879M,2014MNRAS.442.3495V,2023MNRAS.526..512K}. 
Figure \ref{fig:FS} shows that the second break of $\Gamma\beta$ appears at $t\simeq4\times10^6$ s for the thick shell case and $2\times10^6$ s for the thin shell case. These break timescales are almost independent of the initial magnetization of the ejecta. 
Since the second break time scale strongly depends on the initial width of the ejecta, we may infer the radial structure of the ejecta. 
For $R>R_{\rm BM}$, the radial profile of $\Gamma\beta$ agrees with the analytical Blandford-McKee solution (red line in Figure \ref{fig:RF}). The evolution of the physical quantities for $R>R_{\rm BM}$ is the same as the standard model. 

\subsection{Spectrum evolution}

\begin{figure}
\includegraphics[width=\columnwidth]{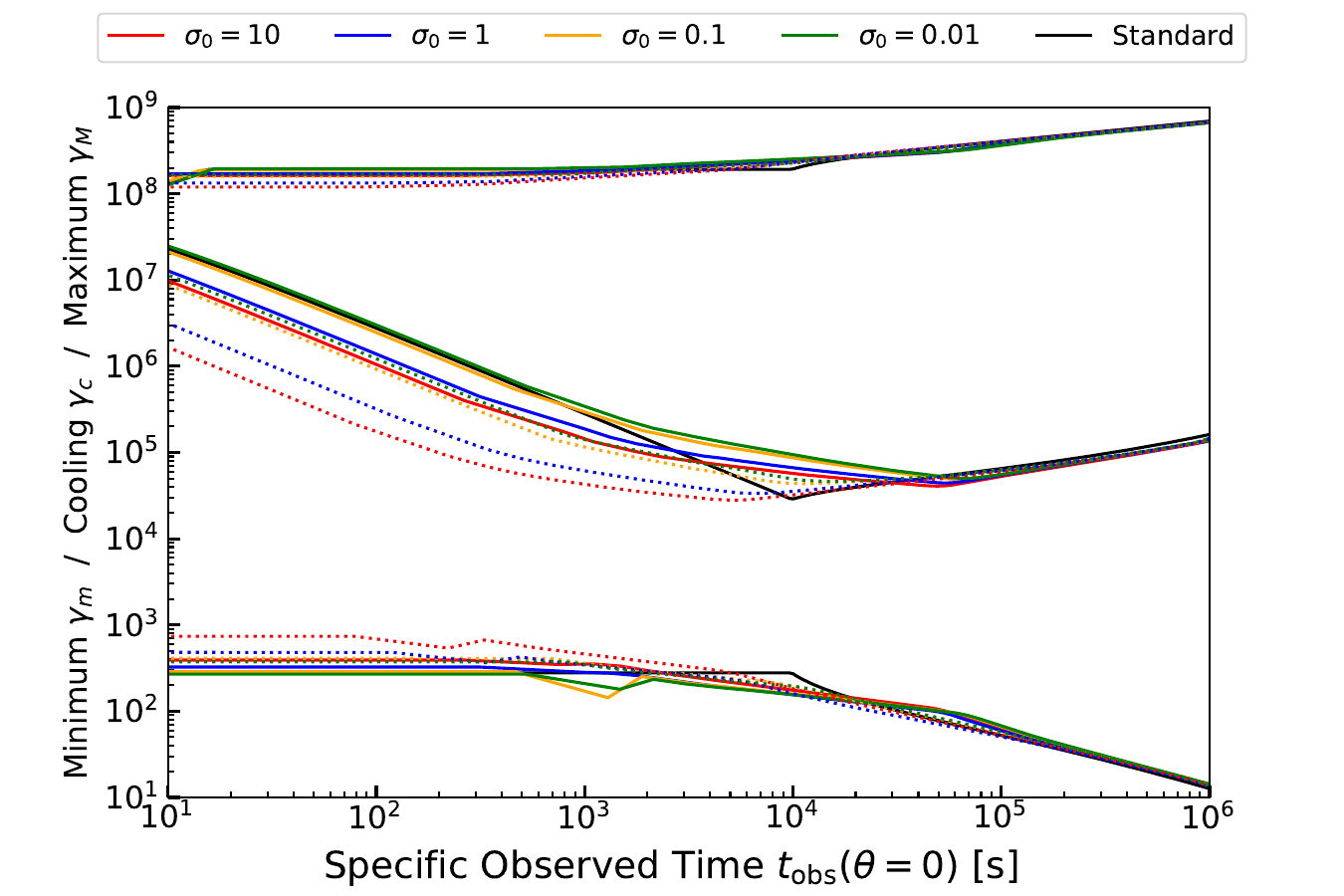}
\caption{The time evolution of the characteristic Lorentz factors of accelerated electrons in the forward shock region. 
From the top, middle, and bottom, we show the maximum, cooling, and minimum Lorentz factor, respectively. The parameters are $E_0=10^{50}$ erg, $\Gamma_0=10$, $n_0=1\ \rm{cm}^{-3}$, $\epsilon_{\rm e}=0.1$, and $\epsilon_{\rm B}=0.01$. The solid and dotted lines are for the thick and thin cases. Each coloured line corresponds to a different initial magnetization $\sigma_0$. The black lines are for the standard model. 
\label{fig:gamma_FS}}
\end{figure}

\begin{figure*}
\includegraphics[width=17.5cm]{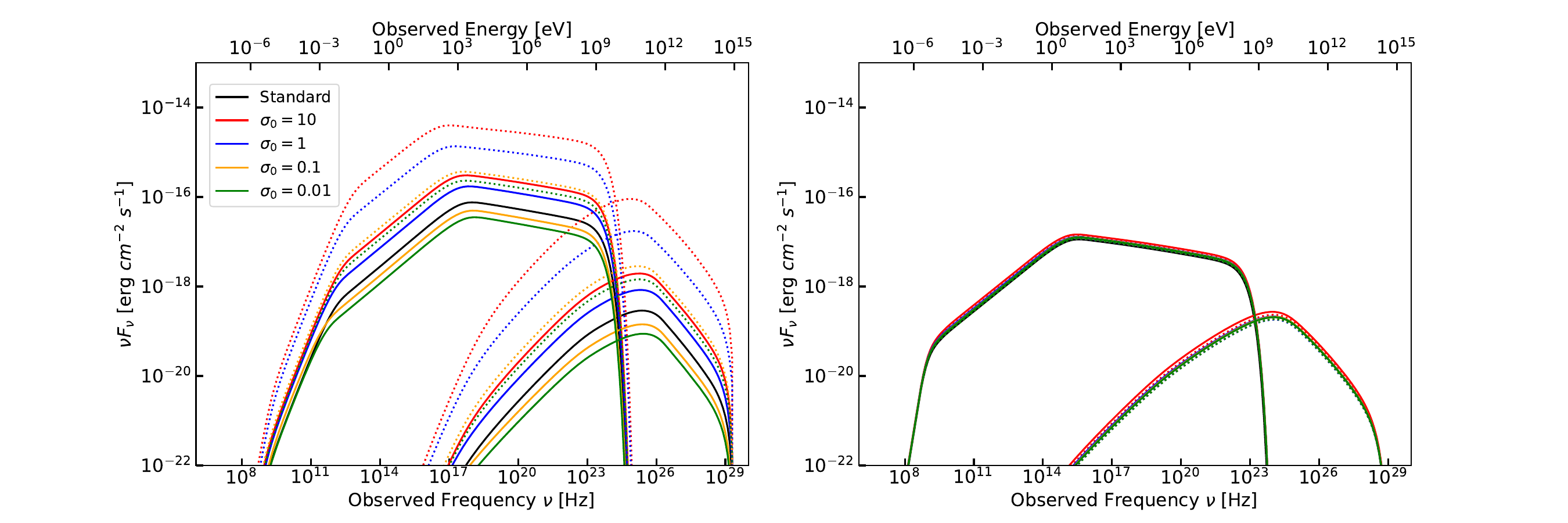}
\caption{The observed spectrum of the forward shock radiation at $t_{\rm obs}=2.0\times10^3$ s. The parameters are $E_0=10^{50}$ erg, $\Gamma_0=10$, $n_0=1\ \rm{cm}^{-3}$, $\epsilon_{\rm e}=0.1$, $\epsilon_{\rm B}=0.01$, and $z=1.0$.
The low-frequency component is the synchrotron radiation and the high-frequency component is the SSC radiation. The solid lines represent the thick shell cases and the dotted ones represent the thin shell cases. Each coloured line corresponds to a different initial magnetization $\sigma_0$. The black line shows the standard model. The right panel: the same figure at $t_{\rm obs}=1.0\times10^6$ s.
\label{fig:spectrum_FS1}}
\end{figure*}

Based on the Lorentz factor evolution of the forward shock, we can calculate the characteristic Lorentz factors of the non-thermal electrons from Eqs. (\ref{eq:gamma_m}), (\ref{eq:gamma_M}), and (\ref{eq:gamma_c}). The result is shown in Figure \ref{fig:gamma_FS}. In our cases, the electron spectra are always slow cooling ($\gamma_{\rm m}<\gamma_{\rm c}$). 
In the high-$\sigma$ case, the high value of the initial Lorentz factor makes the minimum Lorentz factor larger and the cooling Lorentz factor smaller than the standard model. However, these differences gradually disappear in the late phase. 

Then, we produce the corresponding observed radiation spectra as shown in Figure \ref{fig:spectrum_FS1}. The radiation spectra in the transition phase are shown in the left panel, while those in the BM phase are exhibited in the right panel. 
The three spectral breaks in the synchrotron component are characterised by three frequencies: minimum $\nu_{\rm m}\equiv\nu(\gamma_{\rm m})$, cooling $\nu_{\rm c}\equiv\nu(\gamma_{\rm c})$, and maximum frequency $\nu_{\rm M}\equiv\nu(\gamma_{\rm M})$. The spectral slope is roughly followed by the analytical result of \citet{1998ApJ...497L..17S}
\begin{equation}
    \nu F_\nu\propto
    \left\{
    \begin{array}{lll}
    \nu^{4/3}\ \ \ \ \ \ \ \ \ \ \ \ \ \nu<\nu_{\rm m}\\
    \nu^{-\frac{p-3}{2}}\ \ \ \ \ \ \ \nu_{\rm m}<\nu<\nu_{\rm c}\\
    \nu^{-\frac{p-2}{2}}\ \ \ \ \ \ \ \nu_{\rm c}<\nu<\nu_{\rm M}
    \end{array}
    \right. .
    \label{eq:spectra}
\end{equation}

The Klein-Nishina effect makes the SSC component softer than the synchrotron component in the highest energy.
Since the optical photon number density is not so high, the $\gamma\gamma$ annihilation process does not affect the spectrum in our cases. The SSC cutoff energy reflects the maximum accelerated energy of electrons. 

The higher Lorentz factor for the high-$\sigma$ models make the early phase radiation an order of magnitude brighter than the standard model as shown in the left panel of Figure \ref{fig:spectrum_FS1}. In the BM phase, no difference appears among the models as shown in the right panel of Figure \ref{fig:spectrum_FS1}.

\subsection{Multi-wavelength light curve}

\begin{figure*}
\includegraphics[width=17.5cm]{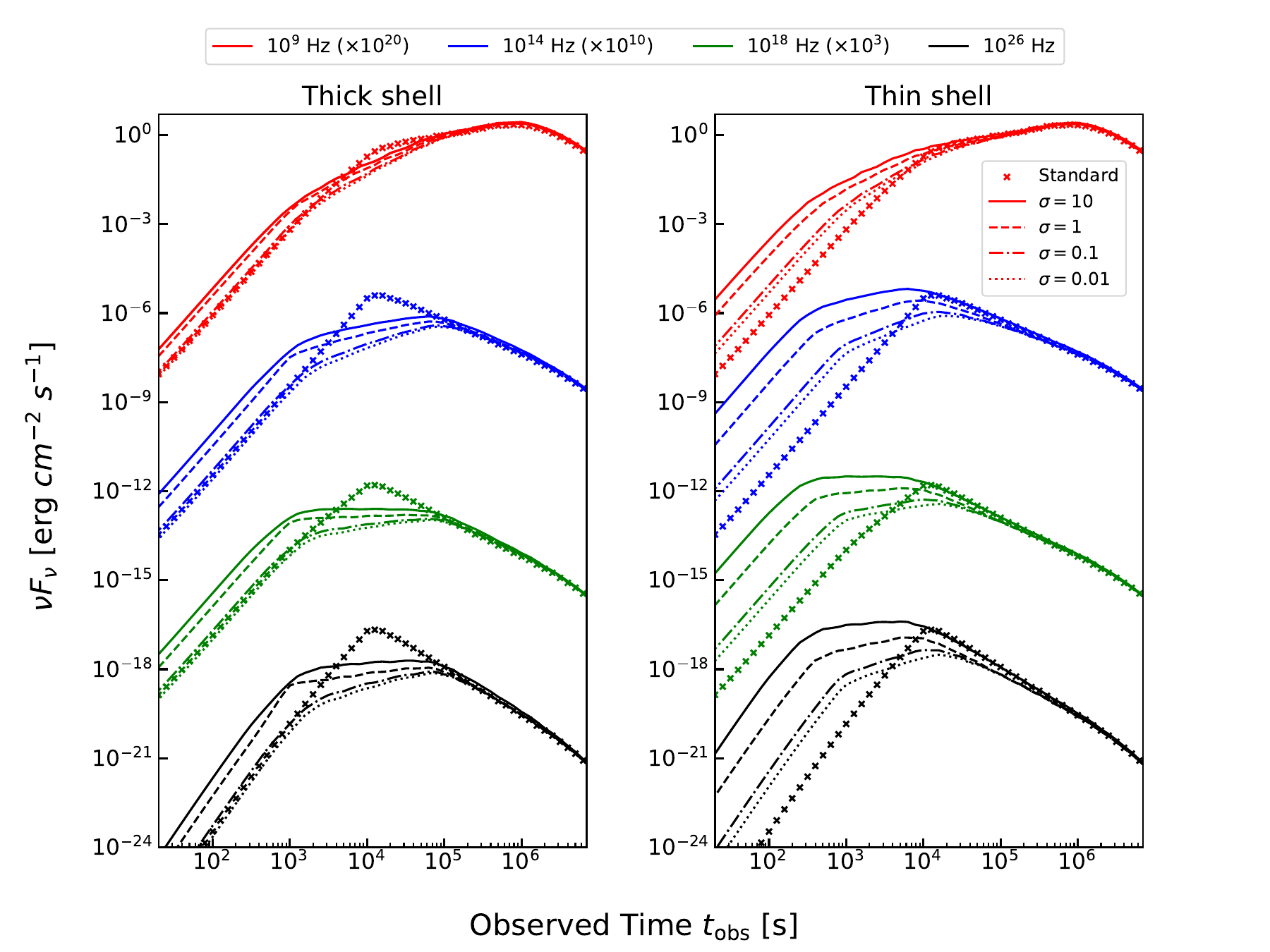}
\caption{The light curves for the case of the forward shock radiation. The left panel shows the thick shell cases while the right panel shows the thin shell cases. The solid, dashed, dash-dotted, dotted lines correspond to the case of $\sigma_0=10,1,0.1,0.01$, respectively. The lines denoted with crosses 
represent the standard model. Each coloured line corresponds to a different observed frequency; red: radio ($10^9$ Hz), blue: optical ($10^{14}$ Hz), green: X-ray ($10^{18}$ Hz), black: high-energy $\gamma$-ray ($10^{26}$ Hz). All plots except for the high-energy $\gamma$-ray have arbitrary offsets for convenience. 
\label{fig:LC_FS}}
\end{figure*}

Figure \ref{fig:LC_FS} shows the light curves for the forward shock radiation. As the late-phase forward shock dynamics are independent of initial magnetization and width, and only depend on the initial energy, the late-phase light curves of all models are consistent with the standard model. The onset time scale for frequencies higher than $\nu_{\rm m}$ corresponds to the beginning of the BM phase depending on the initial width and magnetization of the ejecta. While the onset time scale has been used to estimate the initial bulk Lorentz factor, as our results suggest, the standard model overestimates the Lorentz factor if the initial width is thick. 
For $\nu_{\rm obs}<\nu_{\rm m}$, the flux increases as far as the time when $\nu=\nu_{\rm m}$.

The major difference between our calculations and the standard model is the existence of the transition phase,
which is significantly longer than the reverse shock crossing time. The flux rises gradually compared to that of the coasting phase. The observed flux at the transition phase evolves as $\nu F_\nu\propto t_{\rm obs}^{\beta(\sigma_0)}$, where $\beta(\sigma_0)=0.5\sim2.0$ depending on the initial magnetization. Because this phase lasts until the rarefaction wave catches up timescale, the duration of the transition phase strongly depends on the initial width of the ejecta. These dependencies of the initial width and magnetization are helpful for the investigation of the initial physical conditions of the ejecta. 
\renewcommand{\arraystretch}{1.5}
In the coasting phase, the peak flux follows
\begin{equation}
    \nu F_\nu\propto
    \left\{
    \begin{array}{lll}
    t_{\rm obs}^{2}\ \ \ \ \ \ \ \ \ \ \ \ \ \nu<\nu_{\rm a}\\
    t_{\rm obs}^{3}\ \ \ \ \ \ \ \ \ \ \ \nu_{\rm a}<\nu<\nu_{\rm c}\\
    t_{\rm obs}^{\frac{10-3p}{4-p}}\ \ \ \ \ \ \ \nu_{\rm c}<\nu<\nu_{\rm M}
    \end{array}
    \right. .
    \label{eq:spectra_syn}
\end{equation}
for synchrotron radiation ($\nu_{\rm a}$ is SSA frequency), and 
\begin{equation}
    \nu F_\nu\propto
    \left\{
    \begin{array}{ll}
    t_{\rm obs}^{4}\ \ \ \ \ \ \ \ \ \ \ \ \ \nu<\nu_{\rm c}^{\rm IC}\\
    t_{\rm obs}^{\frac{12-4p}{4-p}}\ \ \ \ \ \ \ \nu_{\rm c}^{\rm IC}<\nu<\nu_{\rm M}^{\rm IC}
    \end{array}
    \right. .
    \label{eq:spectra_SSC}
\end{equation}
for the SSC emission. 
\citep{1999ApJ...520..641S,2001ApJ...548..787S,2005ApJ...619..968W}. 
For the thin shell case of $\sigma_0=10$, the magnetic acceleration boosts the Lorentz factor of the forward shock by factor 3 according to Figure \ref{fig:FS}. From Eq. (\ref{eq:t_obs}), the observed time becomes shorter by a factor of $\Gamma^2\sim10$. If we compare the observed flux at the same observed time, the synchrotron flux of the thin shell case of $\sigma_0=10$ is brighter than the standard model prediction by $\sim10^{3\sim4}$, which is due to the high $\Gamma$ and Doppler boost effects. The SSC flux is also enhanced by $\sim10^{3.5\sim5.5}$ than that of the standard model. Such bright early afterglows can be the best target for future follow-up observations. 

\section{Reverse Shock Emission}\label{sec:RS}

In this section, we study effects of the magnetic acceleration on the dynamics and radiation of the reverse shock. The interaction between the ejecta and the medium generates a reverse shock propagating backwards in the ejecta. The initial kinetic energy of the ejecta is converted into the magnetic and thermal energy of the shocked ejecta. The reverse shock crossing time becomes shorter as the magnetization increases \citep{2005ApJ...628..315Z,2009A&A...494..879M,2023MNRAS.526..512K}.

\subsection{Dynamics of the reverse shock \label{subsec:RS_ISM}}

\begin{figure}
\includegraphics[width=\columnwidth]{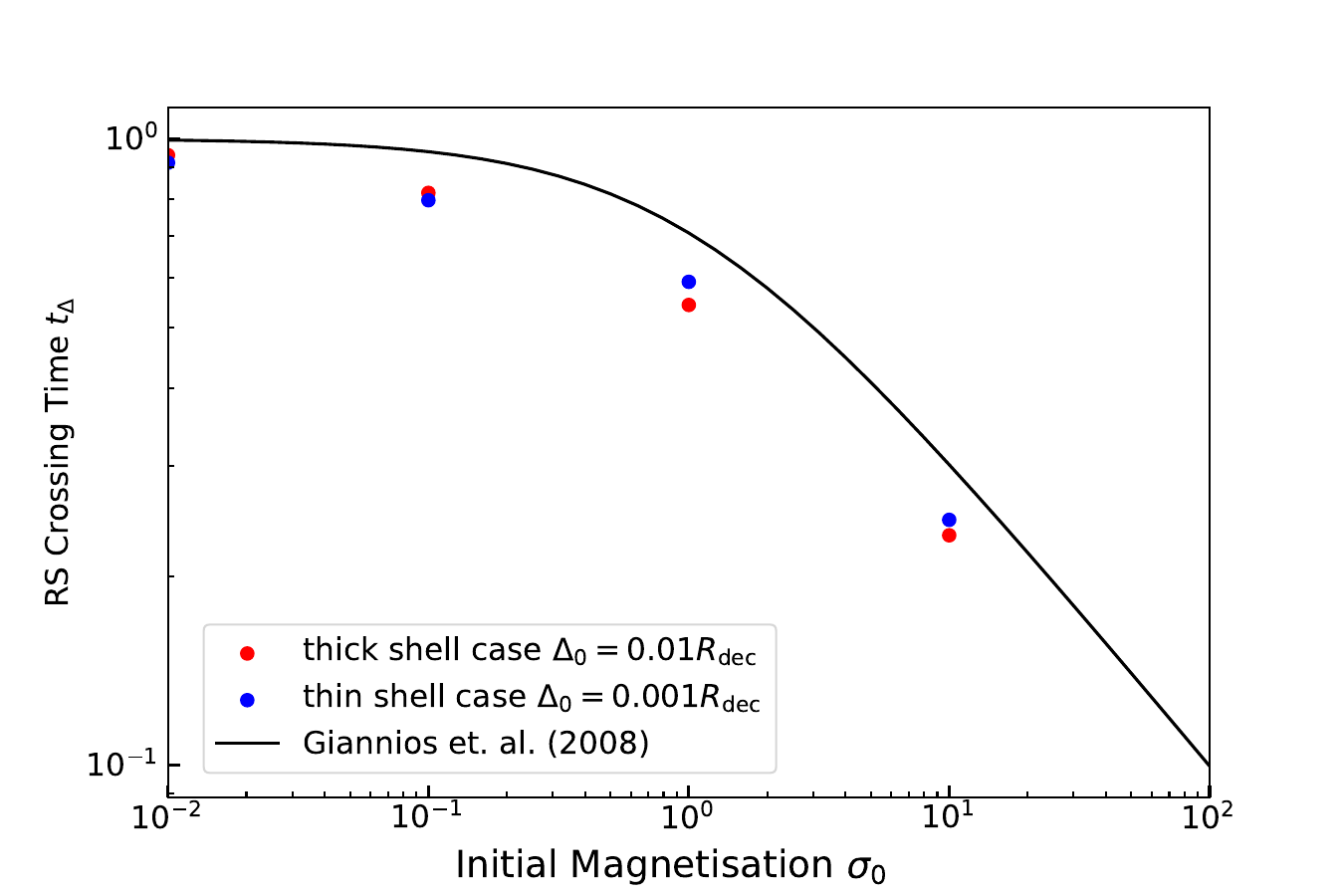}
\caption{The reverse shock crossing timescale as a function of the initial magnetization. The solid line is the analytical estimate given by Eq. (\ref{eq:RS}). 
\label{fig:RS_cross}}
\end{figure}

We define the time at which the reverse shock crosses the ejecta as $t_\Delta$. In a hydrodynamic case ($\sigma_0=0$), \citet{1995ApJ...455L.143S} provided an analytical formula of $t_\Delta$ as
\begin{equation}
    t_\Delta(\sigma_0=0)=\Gamma_0^{1/2}R_{\rm dec}^{3/4}\Delta^{1/4}_0,
    \label{eq:RS_cross}
\end{equation}
where $\Delta_0$ is the shell thickness in the ambient medium rest frame. 
Since a magneto-sonic speed becomes faster in a magnetized medium, the reverse shock crossing time $t_\Delta(\sigma_0)$ becomes shorter \citep{2005ApJ...628..315Z,2008A&A...478..747G} as 
\begin{equation}
\label{eq:RS}
    t_\Delta(\sigma_0)\simeq t_\Delta(\sigma_0=0)(1+\sigma_0)^{-1/2}.
\end{equation}
As shown in Figure \ref{fig:RS_cross}, the above formula well agrees within $\sim$ 10\% with 
numerical simulations \citep[e.g.][]{2009A&A...494..879M,2023MNRAS.526..512K}.
In GRB afterglows, the reverse shock crossing time $t_\Delta(\sigma_0)$ determines the time of the peak flux of the reverse shock radiation \citep{2005ApJ...628..315Z}. Our results confirm that high-$\sigma$ ejecta leads to the early peak time of the reverse shock radiation.

\begin{figure}
\includegraphics[width=\columnwidth]{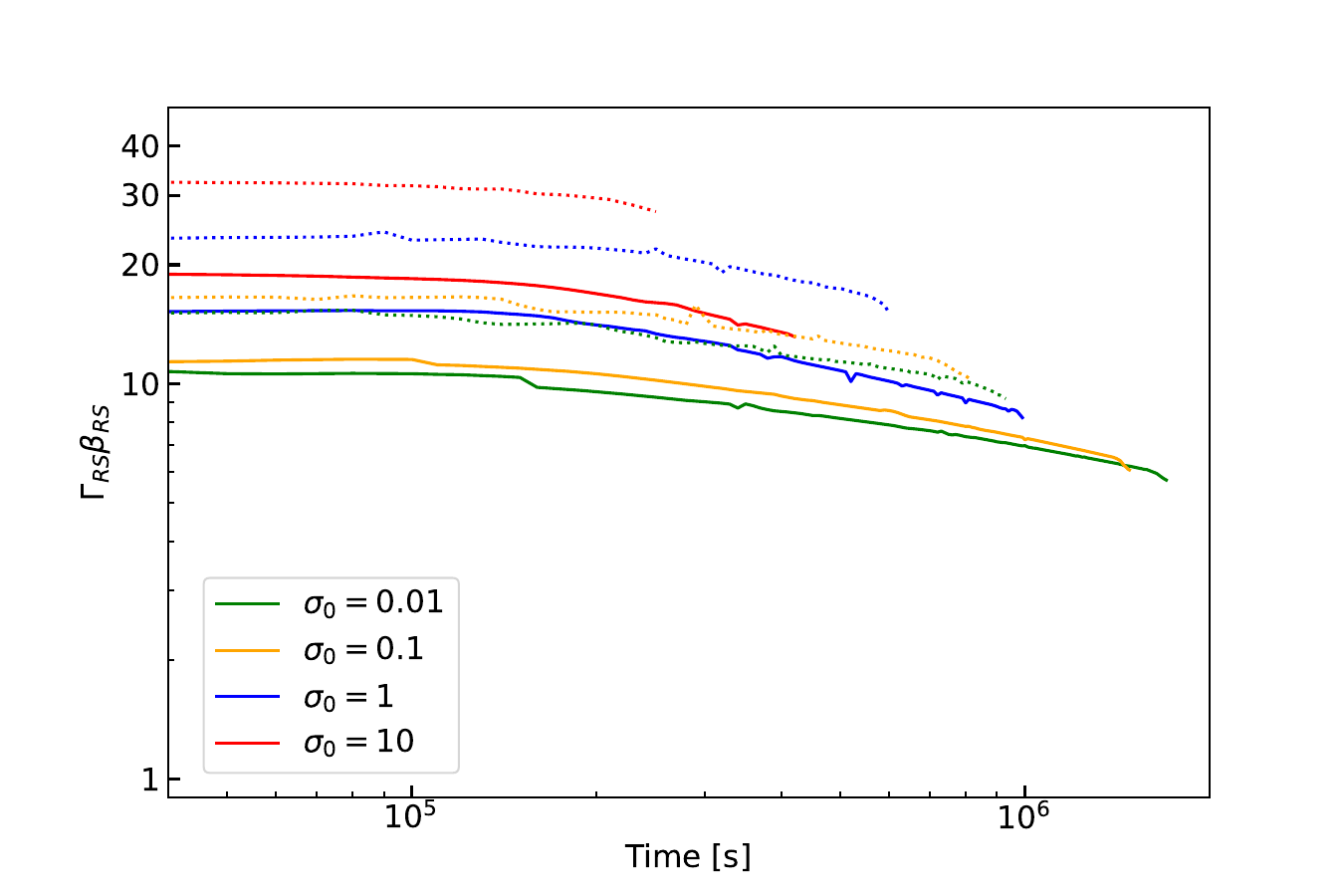}
\caption{The time evolution of $\Gamma\beta$ of the shocked ejecta just behind the reverse shock until the reverse shock crossing time $t_{\Delta}$. The solid line represents the thick shell case and the dotted one represents the thin shell case. Each coloured line corresponds to a different initial magnetization $\sigma_0$. 
\label{fig:RS}}
\end{figure}

Figure \ref{fig:RS} shows the time evolution of $\Gamma\beta$ of the reverse shock, which is estimated at the radius where the Lorentz factor becomes maximum behind the shock front. The data is plotted from a reverse shock ignition time to the crossing time. 
The reverse shock dynamics are almost synchronised with the forward shock dynamics. In high-$\sigma$ cases, the magnetic acceleration boosts the Lorentz factor of the reverse shock. Until the transition phase, the reverse shock speed is almost constant. Then, it decelerates gradually as $t^{-\beta(\sigma_0)}$. The index $\beta(\sigma_0)$ ranges from 0.33 to 0.5 as $\sigma_0 $ increases, which is almost the same value for the case of the forward shock. From our simulation, we confirm the relation $\Gamma_{\rm RS}\simeq\Gamma_{\rm FS}$ \citep{1995ApJ...455L.143S,2005ApJ...628..315Z}. 

Relative Lorentz factor of the reverse shock $\Gamma_{\rm rel}$ is calculated by \citep{1995ApJ...455L.143S,2005ApJ...628..315Z}
\begin{equation}
    \Gamma_{\rm rel}=\Gamma_{\rm RS}\Gamma_{\rm ejecta}(1-\beta_{\rm RS}\beta_{\rm ejecta})\simeq\frac{1}{2}\left(\frac{\Gamma_{\rm ejecta}}{\Gamma_{\rm RS}}+\frac{\Gamma_{\rm RS}}{\Gamma_{\rm ejecta}}\right).
    \label{eq:Gamma_rel}
\end{equation}
In the simulation data, the Lorentz factor 
of the unshocked ejecta $\Gamma_{\rm ejeta}$ is estimated at the radius where the Lorentz factor becomes maximum. Because the speeds are almost the same order $\Gamma_{\rm RS}\sim\Gamma_{\rm ejecta}$ in our simulation results, the strength of the reverse shock is weak $\Gamma_{\rm rel}\sim1$ compared to the forward shock $\Gamma_{\rm FS}\gg1$. {As is well known, the higher magnetization ($\sigma\geqq1$), implying a magneto-sonic speed is $\simeq c$, leads to a weak shock, i.e. a less energy dissipation of the shocked region \citep{1984ApJ...283..694K}. 
Thus, the reverse shock radiation is suppressed for $\sigma\gg1$ \citep{2005ApJ...628..315Z}.} 

\subsection{Multi-wavelength radiation}

\begin{figure}
\includegraphics[width=\columnwidth]{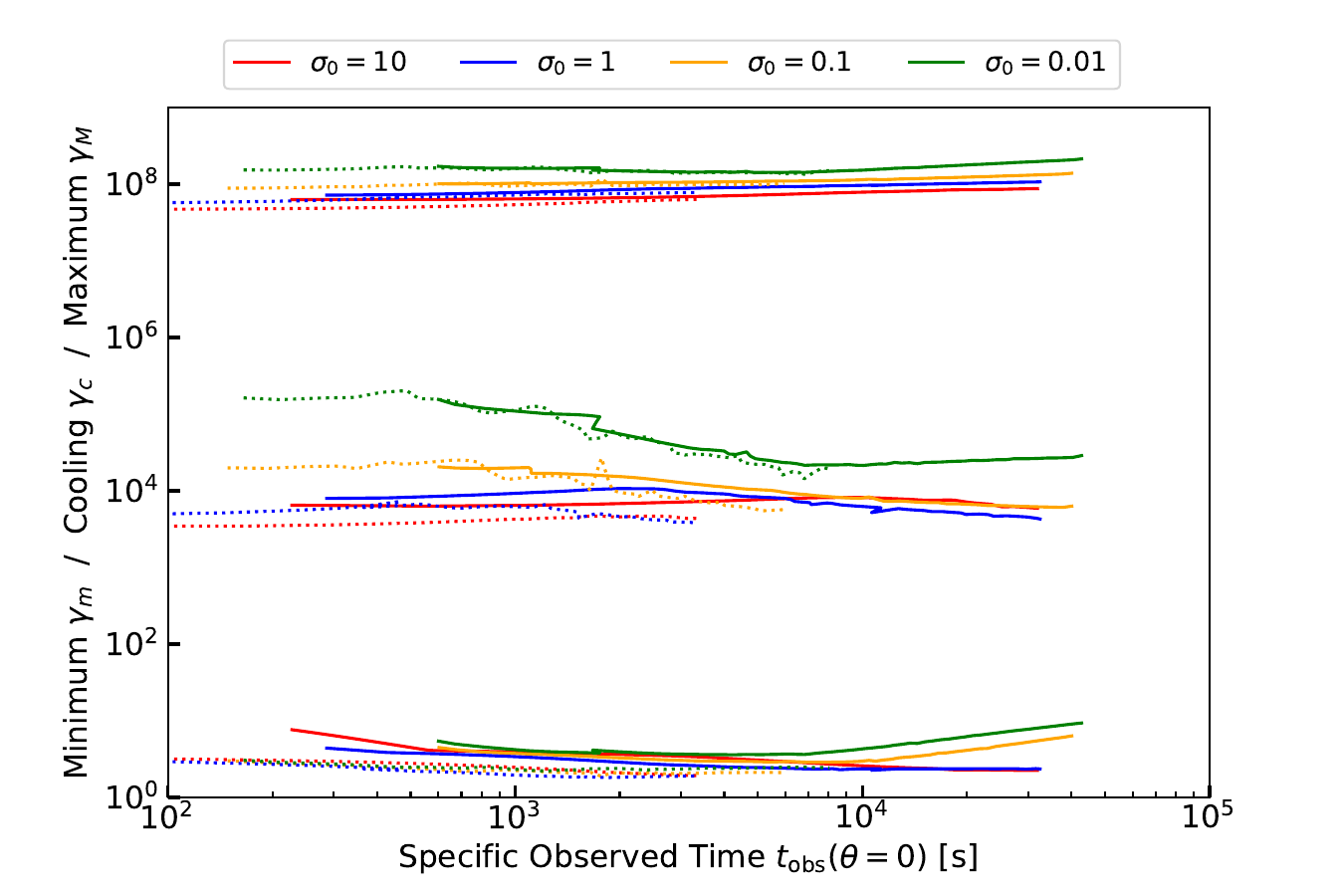}
\caption{The time evolution of the characteristic Lorentz factors of accelerated electrons in the reverse shock region. From the top, middle, and bottom, we show the maximum, cooling, and minimum Lorentz factor, respectively. The parameters are $E_0=10^{50}$ erg, $\Gamma_0=10$, $n_0=1\ \rm{cm}^{-3}$, $\epsilon_{\rm e}=0.1$, and $\epsilon_{\rm B}=0.01$. The solid and dotted lines are for the thick and thin shell cases. Each coloured line corresponds to a different initial magnetization $\sigma_0$. 
\label{fig:gamma_RS}}
\end{figure}

\begin{figure}
\includegraphics[width=\columnwidth]{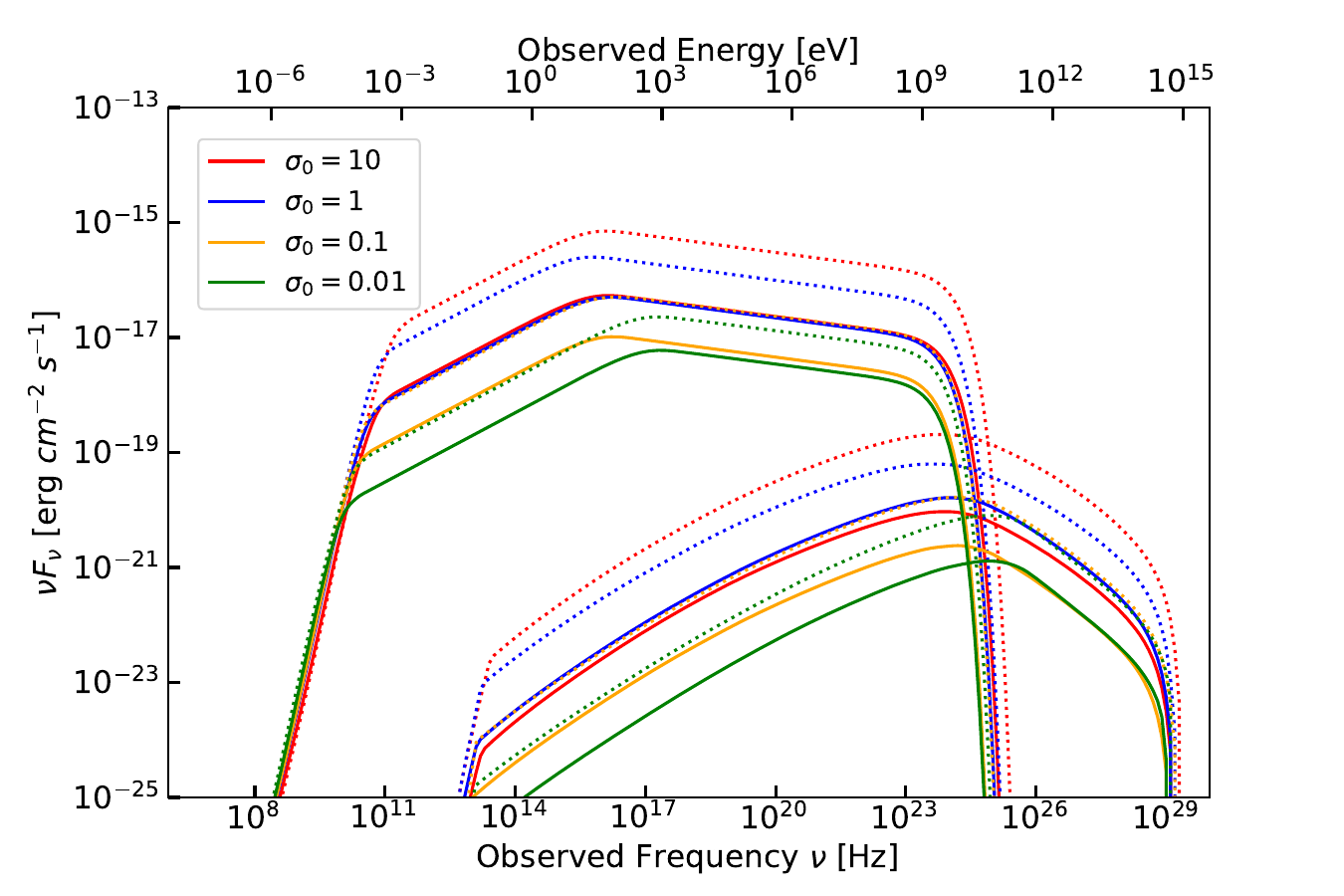}
\caption{The observed spectrum of the reverse shock radiation at $t_{\rm obs}=2.5\times10^3$ s. The low-frequency component is the synchrotron radiation and the high-frequency component is the SSC radiation. The solid lines represent the thick shell cases and the dotted ones represent the thin shell cases. Each coloured line corresponds to a different initial magnetization $\sigma_0$. The black line shows the standard model. 
\label{fig:spectrum_RS1}}
\end{figure}

Based on the Lorentz factor evolution of the reverse shock, we can calculate the characteristic Lorentz factor of non-thermal electrons from Eqs. (\ref{eq:gamma_m}), (\ref{eq:gamma_M}),  and (\ref{eq:gamma_c}). The result is shown in Figure \ref{fig:gamma_RS}. In our simulations, the electron spectra are always slow cooling ($\gamma_{\rm m}<\gamma_{\rm c}$) due to the numerically limited parameter range (especially the initial energy). Since the relative Lorentz factor of the reverse shock is very weak, the minimum Lorentz factor of electrons $\gamma_{\rm m}$ is almost 1 in the coasting phase, then gradually increases in the transition phase for low magnetization cases. 
The cooling Lorentz factor evolves in different ways depending on the initial magnetization. For high magnetization cases, the initial plateau evolution of $\gamma_{\rm c}$ is due to the weakness of the reverse shock compression of the magnetic field. Then, it gradually decreases as ejecta decelerates. For low magnetization cases, on the other hand, relatively strong reverse shock can compress the ejecta magnetic field resulting in the decreasing feature of $\gamma_{\rm c}$. 

The corresponding observed radiation spectra at $2.5\times10^3$ s are shown in Figure \ref{fig:spectrum_RS1}. Due to the relatively high magnetization level compared to the forward shock case, the SSC peak flux is a few orders of magnitude lower than the synchrotron peak flux. The synchrotron spectral shape is roughly followed by Eq. (\ref{eq:spectra}), but hard spectra $\nu F_\nu\propto \nu^{3.5}$ due to SSA is seen in the low-energy region. 
Because $\nu_{\rm m}$ is an order of magnitude smaller than that of the forward shock, the reverse shock emission can emerge in the low-frequency region for low-magnetized cases. 

\begin{figure*}
\includegraphics[width=17.5cm]{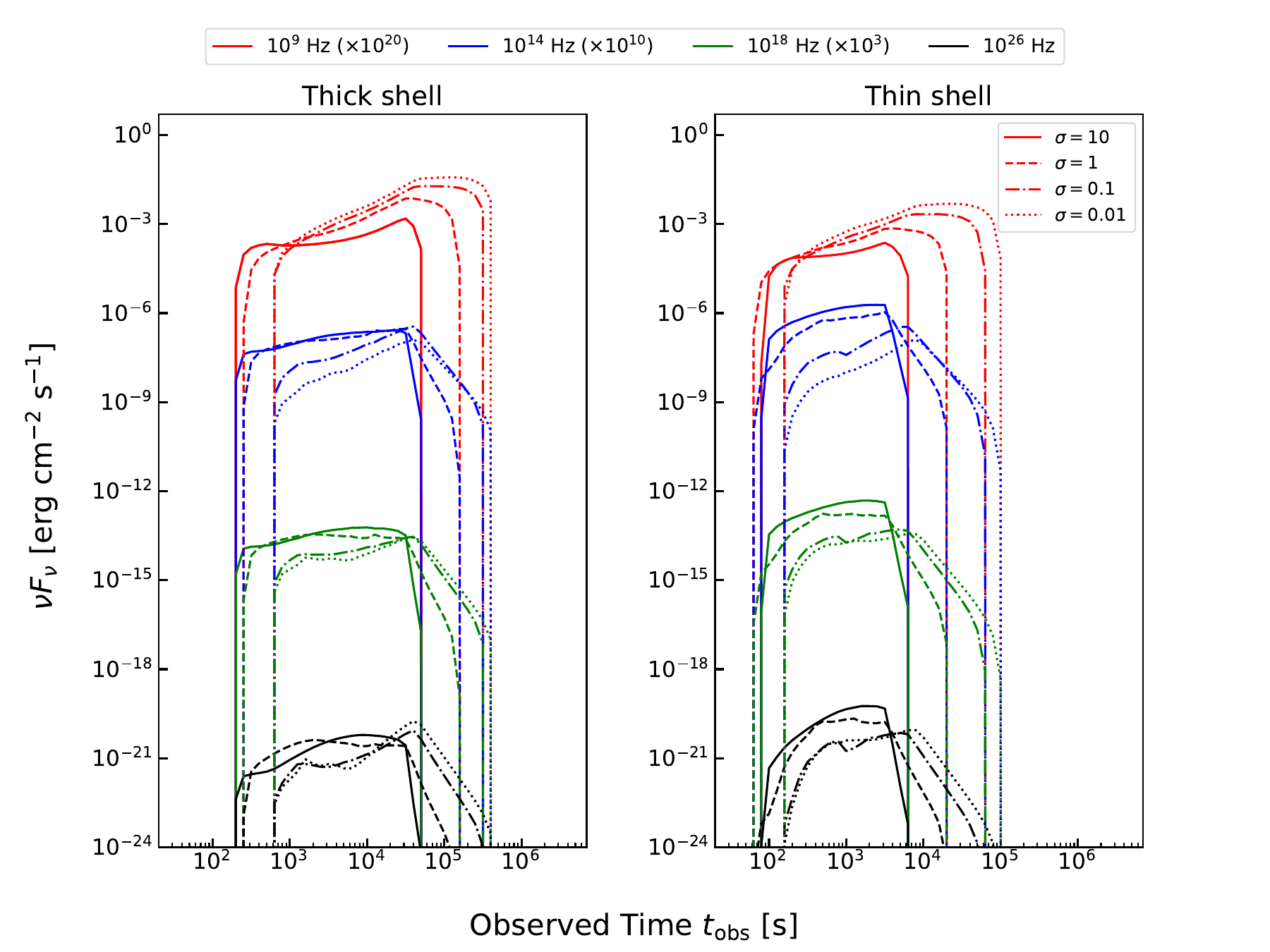}
\caption{The light curves for the case of the reverse shock radiation. The left panel shows the thick shell cases while the right panel shows the thin shell cases. The solid, dashed, dash-dotted, dotted lines correspond to the case of $\sigma_0=10,1,0.1,0.01$, respectively. Each coloured line corresponds to a different observed frequency; red: radio ($10^9$ Hz), blue: optical ($10^{14}$ Hz), green: X-ray ($10^{18}$ Hz), black: high-energy $\gamma$-ray ($10^{26}$ Hz). All plots except for the high-energy $\gamma$-ray have arbitrary offsets for convenience. 
\label{fig:LC_RS}}
\end{figure*}

Figure \ref{fig:LC_RS} shows the light curves for the reverse shock radiation. 
All models achieve the peak flux at the reverse shock crossing time.
Even after the reverse shock crossing time, the high latitude emission lasts for a while, making decreasing features.
Due to the magnetic acceleration, the peak flux of the thin shell cases is slightly higher than that for the thick shell cases. 
The rising slope has a slight dependence on the initial magnetization $\sigma_0$, which might be constrained by early follow-up observations. 
For higher magnetization cases, the weak energy dissipation at the reverse shock suppresses the emission flux. 
Besides the high latitude emission ceases faster for higher magnetization cases due to the short angular timescale $\sim R_{\rm RS}/(c\Gamma^2_{\rm RS})$.

\section{Semi-Analytical Description}\label{sec:model}

\begin{figure}
\includegraphics[width=\columnwidth]{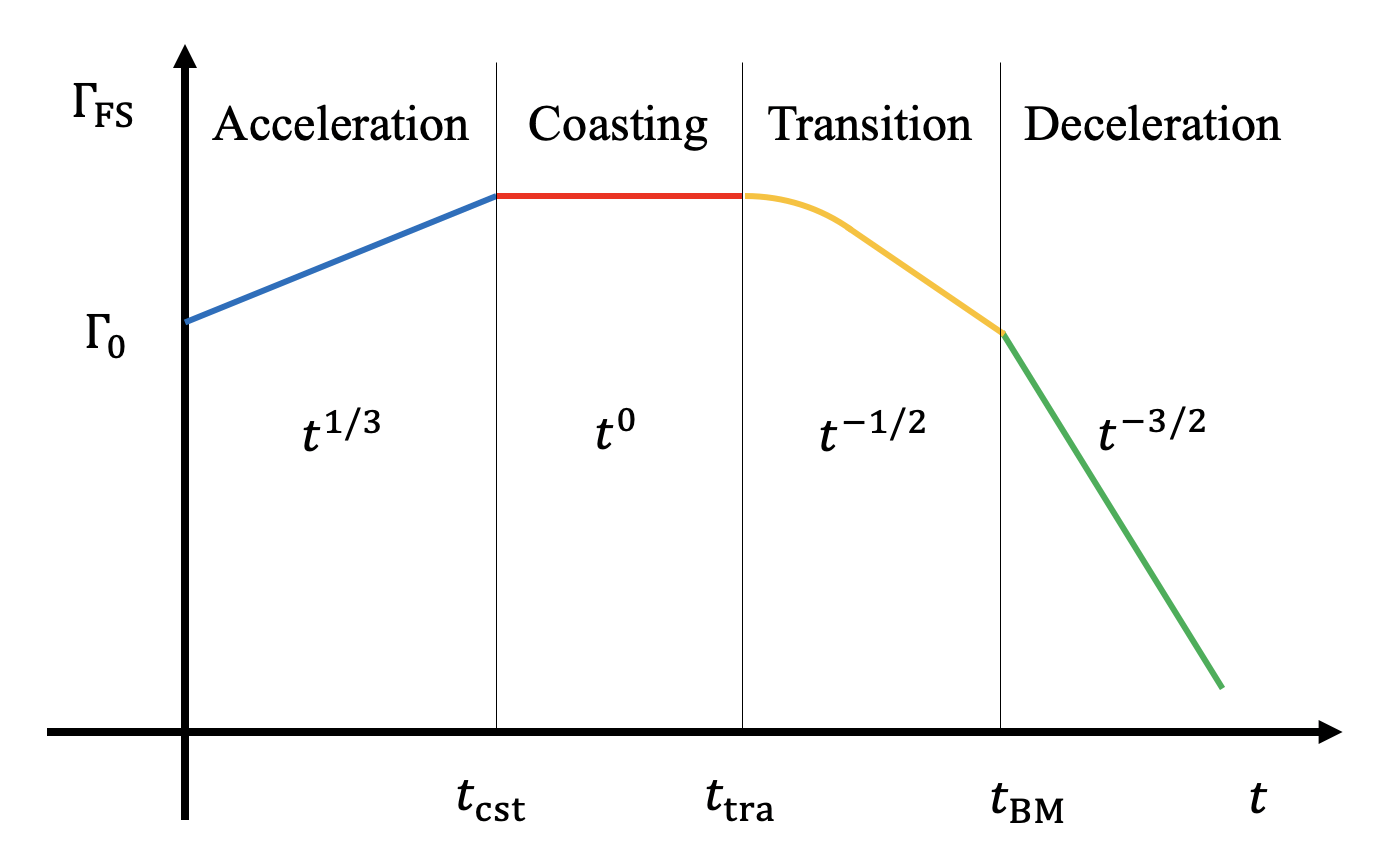}
\caption{A schematic description of the evolution of the forward shock Lorentz factor. Our model consists of 4 phases: acceleration, coasting, transition, and deceleration. 
The timescale $t_{\rm cst}$, $t_{\rm tra}$, and $t_{\rm BM}$ are switching timescales between each phase (see the text in detail). 
\label{fig:Gamma_model}}
\end{figure}

Based on our simulation results, we provide an analytical description of the evolutions of the shocks and emissions. The following formulae include the effect of the initial magnetic acceleration and the transition phase as a function of the initial magnetization and width of the ejecta. Figure \ref{fig:Gamma_model} is a schematic picture of the forward shock dynamics. 
We explain the details and feasibility of our formulae below. 

\subsection{Reverse shock ignition time}
\label{sec:rev}

In high-$\sigma$ cases, the impulsive acceleration of the ejecta is expected \citep{2010PhRvE..82e6305L,2011MNRAS.411.1323G,2012MNRAS.422..326K}. 
Initially, the magnetic pressure in the ejecta is higher than that of the shocked ambient medium. 
When the gas pressure of the shocked ambient medium overwhelms the total pressure of the ejecta, a reverse shock can ignite \citep{2012MNRAS.421.2442G,2012MNRAS.421.2467G}. Since the reverse shock decelerates the ejecta, the acceleration of the forward shock is terminated at the reverse shock ignition time $t_{\rm RS}$.
The pressure balance at the contact discontinuity \citep[e.g.][]{2005ApJ...628..315Z} is written as
\begin{equation}
    \frac{B'^2_{\rm ej}}{8\pi}=\frac{4}{3}\Gamma_{\rm FS}^2n_0m_{\rm p}c^2,
    \label{eq:RS_balance}
\end{equation}
where $B_{\rm ej}$ is the magnetic field of the ejecta. This equation can be written through magnetization as 
\begin{equation}
    \sigma_{\rm ej}=\frac{8}{3}\frac{\Gamma_{\rm FS}^2n_0}{n_{\rm ej}},
\end{equation}
where $\sigma_{\rm ej}$ and $n_{\rm ej}$ are the magnetization and number density of the ejecta, respectively. From number conservation, we can estimate the number density of the ejecta as
\begin{equation}
    n_{\rm ej,0}(t)=\frac{E_0}{4\pi\Delta_0(1+\sigma_0)\Gamma_0^2m_{\rm p}c^2}(R_0+ct_{\rm RS})^{-2}.
    \label{eq:n_ej_0}
\end{equation}

For low magnetization cases ($\sigma_0\leqq1$), we assume $\sigma_{\rm ej}\simeq\sigma_0$.
Even for high magnetization cases ($\sigma_0\geqq1$), a rarefaction wave reduces the magnetic pressure of the ejecta, resulting in an ignition of a reverse shock.
In the impulsive acceleration, the time evolution of the typical Lorentz factor and the magnetization of the ejecta are \citep{2010PhRvE..82e6305L,2011MNRAS.411.1323G}
\begin{equation}
    \Gamma_{\rm ej}(t)=\Gamma_0\left(1+ct\sigma_0/\Delta_0\right)^{1/3},
    \label{eq:imp-G}
\end{equation}
and
\begin{equation}
    \sigma_{\rm ej}(t)=\sigma_0\left(1+ct\sigma_0/\Delta_0\right)^{-1/3},
\end{equation}
respectively. Besides, the number density of the ejecta decreases as
\begin{equation}
    n_{\rm ej}(t)=n_{\rm ej,0}\left(1+ct\sigma_0/\Delta_0\right)^{-1/3}.
    \label{eq:n_impulsive}
\end{equation}
{While the energy-weighted Lorentz factor evolves as discussed above \citep[see also,][]{2012MNRAS.422..326K,2023MNRAS.526..512K}, \citet{2010PhRvE..82e6305L} and \citet{2012MNRAS.421.2467G} claim that the maximum $\Gamma$ of the forward shock is accelerated to $\sim\Gamma_0\sigma_0$ instantaneously for the step-function profile of $\sigma$ expanding into vacuum. 
For predicting the emission property in the actual GRB jets, the approximate method has not been established. In this paper we adopt the energy-weighted behavior as above. To make a semi-analytic model based on our simulation results,} we adopt $\Gamma_{\rm FS}\simeq\Gamma_{\rm ej}$, and then obtain the following fifth-order equation from Eq. (\ref{eq:RS_balance}),
\begin{equation}
    (ct_{\rm RS}+R_0)^3(ct_{\rm RS}+\Delta_0/\sigma_0)^2=C,
\end{equation}
where
\begin{equation}
    C=\sqrt{\frac{\Delta_0}{\sigma_0}\left(\frac{R_{\rm dec}^3}{8\Gamma_0^2(1+\sigma_0)}\right)^3}.
\end{equation}
If a condition $R_0^3(\Delta_0/\sigma_0)^2 \leq C$ is satisfied, there exists only one physical solution for $t_{\rm RS}$ in $0<t_{\rm RS}<C^{1/5}$, otherwise we should take $t_{\rm RS}=0$. 
Adopting the obtained $t_{\rm RS}$ into Eq. (\ref{eq:imp-G}), the saturated Lorentz factor of the ejecta $\Gamma_{\rm sat}$ is estimated as
\begin{equation}
    \Gamma_{\rm sat}=\min(\Gamma_{\rm ej}(t=t_{\rm RS}),\ \Gamma_0(1+\sigma_0)).
    \label{eq:Gamsat}
\end{equation}

\subsection{Shock waves evolutions in the transition phase}
\label{sec:ev-tran}

In the coasting phase, $\Gamma_{\rm FS}=\Gamma_{\rm RS}=\Gamma_{\rm sat}$.
As the ejecta expands, the pressure in the reverse shocked region decreases, but balances with the gas pressure of the shocked ambient medium as \citep{2005ApJ...628..315Z}
\begin{equation}
        (1+\sigma_{\rm RS})(4\Gamma_{\rm rel}+3)(\Gamma_{\rm rel}-1)n_{\rm ej,1} =(4\Gamma_{\rm FS}+3)(\Gamma_{\rm FS}-1)n_0.
\end{equation}
Combining these conditions with Eq. (\ref{eq:Gamma_rel}), we obtain the evolution in the transition phase as
\begin{equation}
    \Gamma_{\rm FS}(t)=\Gamma_{\rm RS}(t)\simeq\frac{\Gamma_{\rm sat}}{\left[ 1+2\Gamma_{\rm sat}\sqrt{\frac{n_0}{n_{\rm ej,1}(1+\sigma_{\rm RS})}} \right]^{1/2}},
    \label{eq:Gamma_trans}
\end{equation}
where $n_{\rm ej,1}$ is the number density of the unshocked ejecta after the end of the impulsive acceleration and $\sigma_{\rm RS}$ is the magnetization of the shocked ejecta. In the limit of the non-magnetized case ($\sigma_0=0$), the expression agrees with the same equation given in \citet{2004MNRAS.353..511P,2007PhRvD..76l3001M,2012ApJ...744...36S}.

In the coasting phase, the second term of the denominator in Eq. (\ref{eq:Gamma_trans}) is negligible.
That term becomes unity at the onset of the transition phase, $t=t_{\rm tra}$. According to our results, the magnetization of the shocked ejecta is roughly the same as the initial magnetization, so we set $\sigma_{\rm RS}\simeq\sigma_0$. The number density of the unshocked ejecta is roughly estimated from {combining Eq. (\ref{eq:n_ej_0}), (\ref{eq:n_impulsive}), (\ref{eq:Gamsat}) as}
\begin{equation}
    n_{\rm ej,1}(t)=\frac{E_0}{4\pi\Delta_0(1+\sigma_0)\Gamma_{\rm sat}\Gamma_0 m_{\rm p}c^2}(R_0+ct)^{-2}.
    \label{eq:nrs}
\end{equation}
Then, we have the following second-order equation {by equating the first and the last terms in the denominator of Eq. (\ref{eq:Gamma_trans})}
\begin{equation}
    (ct_{\rm tra}+R_0)^2\simeq\frac{R_{\rm dec}^3\Gamma_0}{12\Delta_0\Gamma_{\rm sat}^3}.
\end{equation} 
However, the onset time of the transition phase $t_{\rm tra}$ does not well characterise the shock evolution. In the coasting and transition phases, the Lorentz factor evolves following Eq. (\ref{eq:Gamma_trans}), which produces a gradual evolution with Eq. (\ref{eq:nrs}) rather than the sharp break at $t=t_{\rm tra}$.

Since the factor $n_{\rm ej,1}(1+\sigma_{\rm RS})$ decreases as $\propto t^{-2}$,
the asymptotic behaviour of Eq. (\ref{eq:Gamma_trans}) in the transition phase becomes
\begin{equation}
    \Gamma_{\rm FS}\propto t^{-1/2}.
\end{equation}
This index is roughly consistent with our numerical results.
The shocked ejecta sustains the forward shocked region.
Since the energy injection from the shocked ejecta to the forward shock mitigates the forward shock deceleration \citep{2014MNRAS.442.3495V}, the transition phase lasts even after the reverse shock crossing. 

In the acceleration phase, the Lorentz factor $\Gamma_{\rm FS}$ is the same as that of the ejecta $\Gamma_{\rm ej}$ expressed by Eq. (\ref{eq:imp-G}). Then, $\Gamma_{\rm FS}$ gradually deviates from $\Gamma_{\rm ej}$ and transits to the behaviour expressed by Eq. (\ref{eq:Gamma_trans}). Therefore, the maximum value of $\Gamma_{\rm FS}$ is smaller than $\Gamma_{\rm sat}$ given by Eq. (\ref{eq:Gamsat}). To obtain the maximum of $\Gamma_{\rm FS}$, we first define the beginning time of coasting phase $t_{\rm cst}$ by the cross point of the two curves expressed by Eq. (\ref{eq:imp-G}) and Eq. (\ref{eq:Gamma_trans}) as
\begin{equation}
    \Gamma_{\rm ej}(t=t_{\rm cst})=\Gamma_{\rm FS}(t=t_{\rm cst}).
\end{equation}
As $\Gamma_{\rm FS}<\Gamma_{\rm sat}$, the time $t_{\rm cst}$ is earlier than $t_{\rm RS}$.
Adopting the obtained $t_{\rm cst}$ to Eq. (\ref{eq:imp-G}), the maximum value of $\Gamma_{\rm FS}$ is given by
\begin{equation}
    \Gamma_{\rm max}=\Gamma_0(1+ct_{\rm cst}\sigma_0/\Delta_0)^{1/3}.
\end{equation}

\subsection{Transition to BM phase}

The conventional deceleration radius is given by Eq. (\ref{eq:rdec}), which is independent of the initial width of the ejecta. However, our simulations suggest that the transition time to the BM phase $t_{\rm BM}$ is approximately determined by the time at which the rarefaction wave catches up with the forward shock front, which can be roughly obtained from
\begin{equation}
    \Delta=\int_{t_\Delta}^{t_{\rm BM}}(c-v_{\rm FS,front})dt \simeq\int_{t_\Delta}^{t_{\rm BM}}\frac{cdt}{4\Gamma_{\rm FS}^2(t)},
    \label{eq:t_BM}
\end{equation}
where $v_{\rm{FS,front}}=c\sqrt{1-1/(2\Gamma_{\rm FS}^2)}$ is the velocity of the forward shock front, and $\Delta$ is the total width of the shocked ejecta and shocked ambient medium at $t=t_\Delta$ given by Eq. (\ref{eq:RS}).
The time evolution of $\Gamma_{\rm FS}(t)$ is given by Eq. (\ref{eq:Gamma_trans}).
Since the rarefaction wave starts just after the reverse shock crossing time, we can estimate the total width as 
\begin{equation}
    \Delta=\Delta_0+\int_0^{t_\Delta}(v_{\rm FS,front}-v_{\rm FS})dt  \simeq\Delta_0+\int_0^{t_\Delta}\frac{cdt}{4\Gamma_{\rm FS}^2(t)},
\end{equation}
where $v_{\rm FS}$ is the velocity just behind the forward shock. The evolution of $\Gamma_{\rm FS}$ is given by Eq. (\ref{eq:imp-G}) and Eq. (\ref{eq:Gamma_trans}) for $t<t_{\rm cst}$ and $t \geq t_{\rm cst}$, respectively. 
If $\Gamma_{\rm FS}(t_\Delta)$ is larger than $\Gamma_0$ through magnetic acceleration or $\Delta_0$ is sufficiently long, the onset time scale can be longer than the conventional deceleration time $R_{\rm dec}/c$. 
This elongated transition phase is demonstrated by our simulated light curves as shown in Fig. \ref{fig:LC_FS}.

\subsection{Comparison with our simulation and model}

\begin{figure}
\includegraphics[width=\columnwidth]{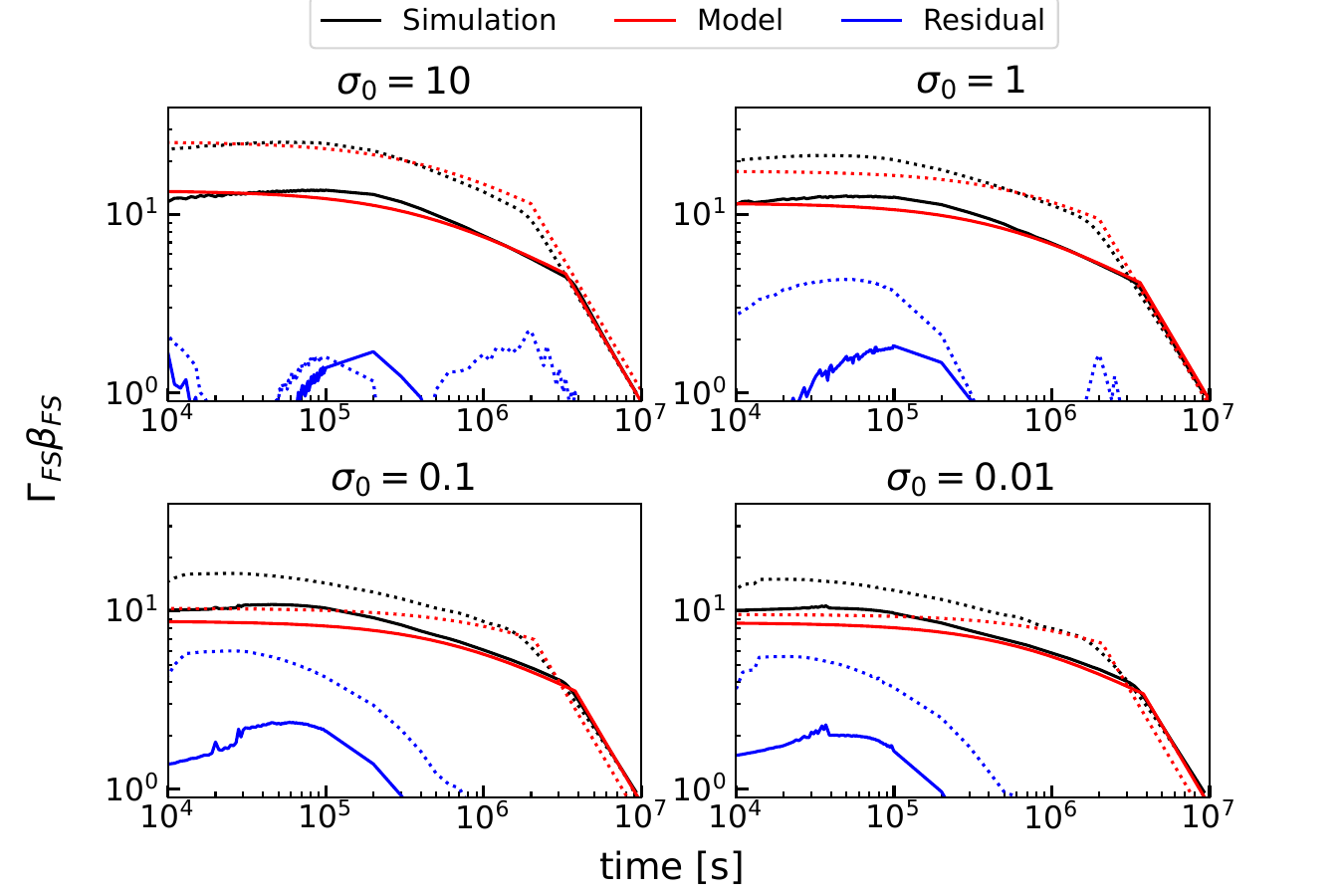}
\caption{The evolutions of the forward shock Lorentz factors obtained by our semi-analytic estimate (red), numerical simulations (black), {and their absolute residuals (blue)}. Each panel shows the dynamics with a different initial magnetization. The solid line corresponds to the thick shell case, while the dotted line corresponds to the thin shell case. 
\label{fig:Gamma_comp}}
\end{figure}

Following the above procedures, we reproduce the dynamics of the forward shock using the same initial parameters in our numerical simulations. The results are shown in Figure \ref{fig:Gamma_comp}. Although some slight differences appear in the early phase of the expansion, our semi-analytic model is almost consistent with our numerical simulation results. The differences are partially due to the finite temperature and finite resolution in our simulations. The thermal pressure expansion slightly accelerates the ejecta even for low magnetized cases. The forward shock is resolved with a few cells, which makes the Lorentz factor small compared to our model. In the late phase dynamics is well described by our model, which means that the afterglow peak time is indeed determined by the rarefaction wave catch-up time.

\section{Application to GRB Early Afterglows}\label{sec:afterglow}

Some GRB early afterglows have mysterious features - optical plateau, optical shallow rising, X-ray shallow decay, gamma-ray very steep rising, and so on \citep{2006MNRAS.369..197F,2009MNRAS.395..490O,2014MNRAS.442.3495V,2018pgrb.book.....Z,2023Sci...380.1390L}. 
Some of those features may be due to the non-monotonic evolution of the bulk Lorentz factor as shown in the previous sections. 

\begin{figure*}
\includegraphics[width=17.5cm]{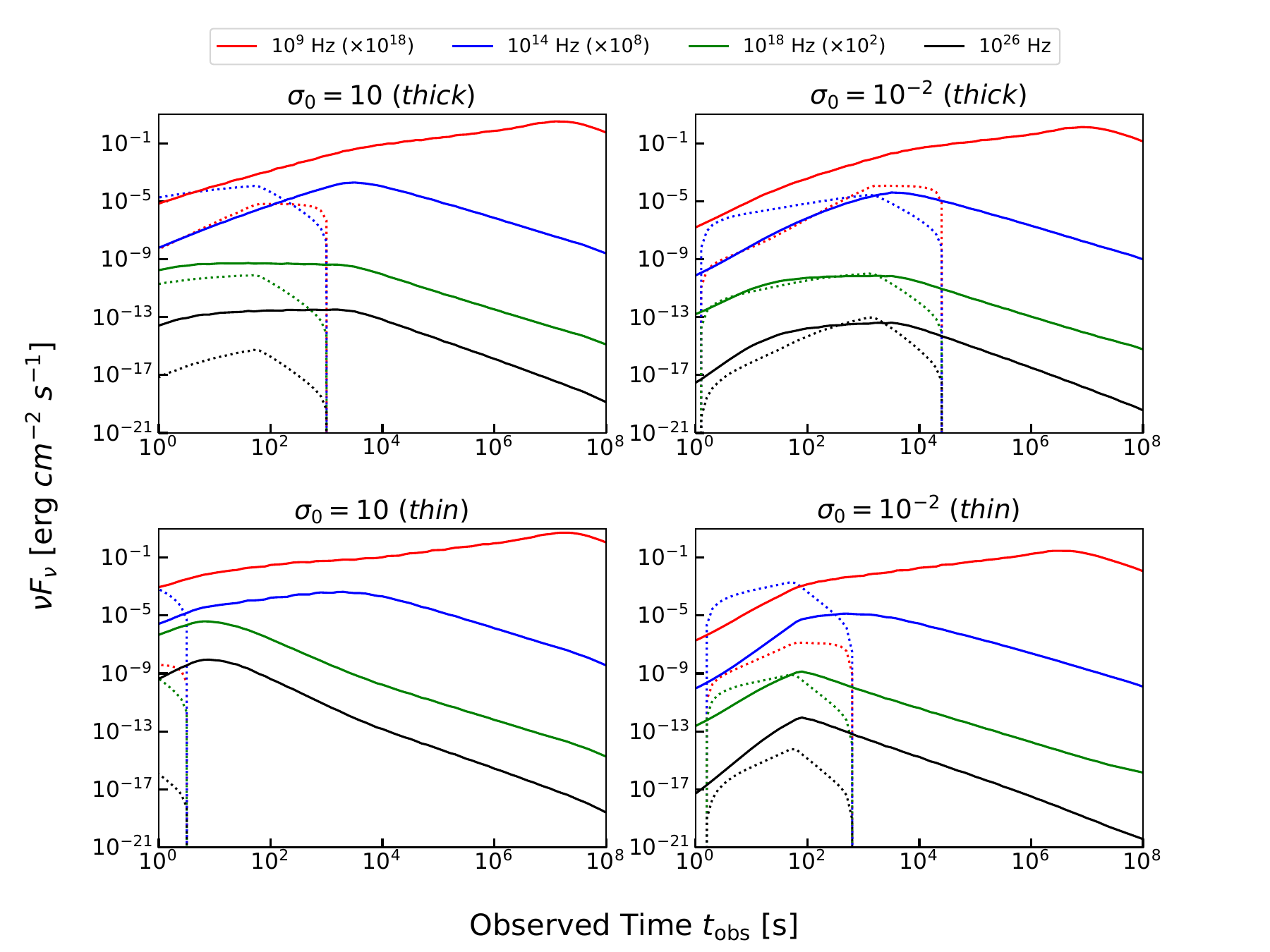}
\caption{The observed light curve combined with the forward and reverse shock radiation. Each panel corresponds to the different initial magnetization and width of the ejecta. The rest of the parameters are the same for all models: $E_0=10^{52}$ erg, $\Gamma_0=100$, and $n_0=1\ \rm{cm}^{-3}$. 
The solid line is the forward shock lightcurve and the dotted one is the reverse shock lightcurve.
Each coloured line corresponds to a different observed frequency; red: radio ($10^9$ Hz), blue: optical ($10^{14}$ Hz), green: X-ray ($10^{18}$ Hz), black: $\gamma$-ray ($10^{26}$ Hz). All data points except for the high-energy $\gamma$-ray have arbitrary offsets for convenience. 
\label{fig:LC_total}}
\end{figure*}

In this section, we discuss the observational implications for thick ($\Delta_0=4R_{\rm dec}/\Gamma_0^2$) and thin shell ($\Delta_0=R_0/\Gamma_0^2$) cases, with $\sigma_0=10^{-2}$ and $\sigma_0=10$. 
We fix the initial ejecta energy $E_0=10^{52}$ erg, initial Lorentz factor $\Gamma_0=100$, initial radius $R_0=R_{\rm dec}/\Gamma_0$, and ambient number density $n_0=1\ \rm{cm}^{-3}$ with $\epsilon_{\rm e}=0.1$ for both the forward and reverse shocks. {The value of $\epsilon_{\rm e}$ in the forward shock is generally different from that of the reverse shock \citep[e.g.][]{2005ApJ...628..315Z,2020ApJ...896..166R}, but we choose the same value for simplicity.} For the reverse shock, we estimate the turbulent magnetic field from the magnetization and the number density of the reverse shock. Meanwhile, $\epsilon_B=0.01$ is adopted for the forward shock. 
In these parameter sets, the shell width gives the timescale $\Delta_0/c=7.2\times10^2$ s for thick shell cases and $\Delta_0/c=1.8$ s for thin shell cases. The other characteristic timescales are summarised in Table \ref{table:models}. 
For the low-magnetized thin shell case, the transition time is longer than the deceleration time $t_{\rm tra}>t_{\rm BM}$. No transition and acceleration phases appear in this case, hence the Lorentz factor evolution in this case is the same as the standard model.
Even for $\Gamma_{\rm FS}>100$, given the evolution of the Lorentz factor calculated with the method in Section \ref{sec:model}, we can produce model light curves of forward and reverse shock radiation following the same procedure written in Section \ref{sec:radiation}. The results are shown in Figure \ref{fig:LC_total}. Due to the large Lorentz factor, the acceleration phase cannot be seen in $t_{\rm obs}>1$ s. 

From our calculated light curves, we can estimate the coefficient $\kappa$ connecting between $t$ and $t_{\rm obs}$;
\begin{equation}
    t_{\rm obs}=(1+z)\frac{t}{\kappa\Gamma^2_{\rm FS}}.
\end{equation}
At $t=t_{\rm BM}$, we can find $\kappa=4$ is the best fitting value for all the cases. As we have mentioned, a light curve break at $t=t_{\rm tra}$ is hard to identify. At $t=t_{\rm cst}$, we find $\kappa=1$ for $\sigma_0=10$ cases (no acceleration phase for $\sigma_0=10^{-2}$).
While $\kappa=4$ is frequently used \citep{1997ApJ...489L..37S,1998ApJ...497L..17S,1999ApJ...519L..17S} to produce light curve analytically, the coefficient $\kappa$ is not always universal \citep[e.g.][]{2021ApJ...923..135D,2024ApJ...970..141A}. 

\begin{table*}
 \caption{Characteristic timescales and the maximum Lorentz factor in Section \ref{sec:afterglow}. }
 \label{table:models}
 \centering
  \begin{tabular}{ccccccc}
   \hline
   Model & $t_{\rm cst}$ [s] & $t_{\rm RS}$ [s] & $t_{\rm tra}$ [s] & $t_{\Delta}$ [s] & $t_{\rm BM}$ [s] & $\Gamma_{\rm max}$ \\
   \hline \hline
   thick ($\sigma_0=10$) & $2.0\times10^3$ & $3.7\times10^3$ & $1.7\times10^4$ & $7.7\times10^5$ & $6.3\times10^6$ & $306$ \\
   thick ($\sigma_0=10^{-2}$) & - & $1.1\times10^4$ & $2.4\times10^5$ & $2.5\times10^6$ & $5.4\times10^6$ & $100$ \\
   \hline
   thin ($\sigma_0=10$) & $2.2\times10^2$ & $1.0\times10^3$ & $1.2\times10^5$ & $1.7\times10^5$ & $1.6\times10^6$ & $1040$ \\
   thin ($\sigma_0=10^{-2}$) & - & $1.5\times10^4$ & $5.1\times10^6$ & $5.7\times10^5$ & $1.2\times10^6$ & $100$ \\
   \hline
  \end{tabular}
\end{table*}

\subsection{Reverse shock radiation}

Candidates of reverse shock components have been observed in radio to optical band from early afterglows, {including the famous examples such as GRB 990123 \citep{1999ApJ...517L.109S}, GRB 021004 \citep{2003ApJ...582L..75K}, GRB 080319B \citep{2018ApJ...859...70F},  GRB 130427A \citep{2013ApJ...776..119L}, and GRB 190829A \citep{2022MNRAS.512.2337D}.}

The reverse shock radiation plotted as the dotted line in Figure \ref{fig:LC_total} is calculated from $t_{\rm RS}$ to $t_\Delta$.
We calculate the width of the reverse shocked region in the shocked region rest frame as 
\begin{equation}
    \Delta R(t)=\Gamma_{\rm RS}(t)\int_0^t(\beta_{\rm ej}-\beta_{\rm RS})cdt,
\end{equation}
where $\beta_{\rm RS}$ is the velocity just behind the reverse shock in the engine-rest frame.
The number and internal energy density of the reverse shocked region are calculated from Eqs. (\ref{eq:Gamma_rel}), and (\ref{eq:nrs}) as
\begin{flalign}
    &n_{\rm RS}=(4\Gamma_{\rm rel}+3)n_{\rm ej,1}, \\
    &\epsilon_{\rm RS}=(\Gamma_{\rm rel}-1)n_{\rm RS},
\end{flalign} 
respectively.
Since the initial magnetization is large and the $\Gamma_{\rm rel}\sim1$, the magnetic field can be roughly estimated as 
\begin{equation}
    B_{\rm RS}'=\sqrt{4\pi n_{\rm RS}m_{\rm p}c^2\sigma_0}.
\end{equation} 
If the initial magnetization is significantly small as $\sigma_0\ll1$, we should estimate $B'_{\rm RS}$ from Eq. (\ref{eq:epsilon_B}) with $\epsilon_B$.  
In the reverse shock light curves, the peak time corresponds to $t_\Delta$.
Even after the peak, the high latitude emission lasts, producing decreasing features. For a thicker shell with a lower magnetization,
the reverse shock ignition time determined by $t_{\rm RS}$ becomes earlier, and
the end of the reverse shock emission becomes longer. 

For high magnetization cases, the shock energy dissipation is inefficient producing the weak emission, and vice-versa. 
As shown in the figures, the reverse and forward shock components can be comparable in optical, X-ray, and gamma-ray bands. The identification of the reverse shock emission in the lightcurve data is key to determining the magnetization.
A caveat is that we assume efficient particle acceleration even in the high magnetization cases, though the diffusive shock acceleration may be inefficient for $\sigma>10^{-3}$ \citep{2011ApJ...726...75S,2013ApJ...771...54S,2018MNRAS.477.5238P}. 
{However, the magnetic reconnection may have crucial roles for particle acceleration in high magnetization cases \citep{2014ApJ...783L..21S,2015MNRAS.450..183S,2019ApJ...880...37P}. The polarization measurements for reveres shock components may have the potential to identify the acceleration mechanism.}
{The reverse shock radiation also depends on the width of the ejecta. As shown in Figure \ref{fig:LC_total}, thin shell cases generate relatively short and bright reverse shock components in optical bands. Such components may be detected as optical flash like GRB 130427A \citep{2014Sci...343...38V}. For thick shell cases, the contribution of both forward and reverse shock components may produce double peak or plateau in optical bands, which can be a key to understanding the multi-wavelength behaviour of the shallow decay phase.}

\subsection{Onset time and flat light curve in the transition phase} \label{subsec:shallow-decay}

The onset timescale has been theoretically thought to be determined by the traditional deceleration timescale for $\nu>\nu_{\rm m}$ (for $\nu<\nu_{\rm m}$, however, a chromatic peak appears at the $\nu_{\rm m}$ crossing timescale) {\citep{2007ApJ...655..973K,2023ApJ...942...34R}. }
The deceleration length scale given by Eq. (\ref{eq:rdec}) provides
the observed deceleration time as
\begin{equation}
\begin{split}
    t_{\rm dec}&\simeq(1+z)\frac{R_{\rm dec}}{2c\Gamma^2_0}\\
    &\simeq10^2\ \rm{s}\ (1+z)\left(\frac{E_0}{10^{52}\ \rm{erg}\ \rm{s}^{-1}}\right)^{1/3}\left(\frac{n_0}{1\ \rm{cm}^{-3}}\right)^{-1/3}\left(\frac{\Gamma_0}{100}\right)^{-8/3},
\end{split}
\end{equation}
where we use typical values for GRB afterglows.
As Figure \ref{fig:LC_total} suggests, even for the same $R_{\rm dec}$, the onset time has diversity within $O(10^1)\sim O(10^3)$ s. For the thin shell case with $\sigma_0=10^{-2}$, the onset time is consistent with the conventional estimate.
On the other hand, the thin shell with $\sigma_0=10$ experiences efficient magnetic acceleration, leading to a shorter onset time due to a larger Lorentz factor than $\Gamma_0$. For the thick shell cases, the peak time is affected by the transition phase. {The existence of the transition phase is basically determined by the thickness of the ejecta: $\Delta_0 > R_{\rm dec}/(\Gamma_0^2c)$.}

The transition phase produces a flatter/rising slope \citep{2009A&A...494..879M,2014MNRAS.442.3495V} compared to the lightcurve at the coasting phase $F_\nu\propto t_{\rm obs}^{2-3}$ in the conventional model \citep{1999ApJ...520..641S}. 
Since the transition phase lasts until the rarefaction catch-up time $t_{\rm BM}$, the thick shell models are preferred to make a long-lasting flat lightcurve as shown in the top panels of Figure \ref{fig:LC_total}. 
The Lorentz factor in the transition phase decreases as $\Gamma_{\rm FS}\propto t^{-1/2}\propto t_{\rm obs}^{-1/4}$. Then, we find the time evolution of the observed flux as
\begin{equation}
    \nu F_\nu\propto
    \left\{
    \begin{array}{lll}
    t_{\rm obs}^{4/3} \ \ \ \ \ \ \ \ \ \ \ \nu<\nu_{\rm m}\\
    t_{\rm obs}^{-\frac{p-3}{2}}\ \ \ \ \ \ \ \nu_{\rm m}<\nu<\nu_{\rm c}\\
    t_{\rm obs}^{-\frac{p-2}{2}} \ \ \ \ \ \ \ \nu_{\rm c}<\nu<\nu_{\rm M}
    \end{array}
    \right. ,
\end{equation}
for the slow cooling case ($\nu_{\rm m}<\nu_{\rm c}$), and
\begin{equation}
    \nu F_\nu\propto
    \left\{
    \begin{array}{lll}
    t_{\rm obs}^{4/3} \ \ \ \ \ \ \ \ \ \ \ \nu<\nu_{\rm c}\\
    t_{\rm obs}^{1/2} \ \ \ \ \ \ \ \ \ \ \ \nu_{\rm c}<\nu<\nu_{\rm m}\\
    t_{\rm obs}^{-\frac{p-2}{2}}\ \ \ \ \ \ \ \nu_{\rm m}<\nu<\nu_{\rm M}
    \end{array}
    \right.,
\end{equation}
for the fast cooling case ($\nu_{\rm c}<\nu_{\rm m}$).
A gradual flux evolution $F_\nu\propto t_{\rm obs}^{1.1}$ in optical bands was observed in GRB 080710A \citep{2009A&A...508..593K}. While this index cannot be explained by standard on and off-axis structured jet scenario \citep{2024JHEAp..41....1O}, our model expectation $F_\nu\propto t_{\rm obs}^{4/3}$ is well consistent with it. 

Assuming the typical particle spectral index of $p=2.2$, the transition phase can produce plateau lightcurve for high frequency (slow cooling: $t_{\rm obs}^{0.4},\ t_{\rm obs}^{-0.1}$; fast cooling: $t_{\rm obs}^{-0.1}$). Such a plateau emission was observed as optical plateau: GRB 050801 \citep{2006ApJ...638L...5R}, XRF 071031 \citep{2009ApJ...697..758K}, GRB 080310 \citep{2012MNRAS.421.2692L}, GRB 080603A \citep{2011MNRAS.417.2124G}, and also as enigmatic X-ray shallow decay phase observed in most GRBs \citep{2019ApJS..245....1T,2019ApJ...883...97Z}  \citep[but not so evident for GeV/TeV-detected GRBs][]{2020MNRAS.494.5259Y}. {Our model also expects GeV/TeV plateau emission during the shallow decay phase, which can be tested by future observations.}

The end of the shallow decay phase is determined by the rarefaction catch-up times $t_{\rm BM}$ given in Eq. (\ref{eq:t_BM}), which strongly depends on the initial width of the ejecta. 
Although the timescale ($\sim 700$ s) corresponds to the shell width in our thick shell model is slightly longer than the typical duration of the prompt emission, the flat lightcurves are consistent with the duration (a few $10^3$ s) of the shallow decay phase. Our model is similar to the energy injection model, {which requires long-lived central engine activity such as late fallback accretion or new-born magnetar} \citep{2001ApJ...552L..35Z,2006ApJ...642..354Z,2006MNRAS.366L..13G,2024ApJ...970..141A}; In our case, however, the effective energy injection from the shocked ejecta lasts for a significantly longer timescale than the reverse shock crossing time. We do not need a long reverse shock crossing time close to the shallow decay duration as assumed in \citet{2014MNRAS.437.2448L}. Furthermore, to reproduce the shallow decay, we do not need to adjust the evolution of the energy injection {from the central engine} as assumed in the conventional model. 

\subsection{TeV afterglow in the acceleration phase}

Because the acceleration phase is very short for observers, the emission in this phase is difficult to detect. However, if the initial Lorentz factor is small and the width is sufficiently thin, it can be detected at most $\sim10$ s. The Lorentz factor in the acceleration phase increases as $\Gamma_{\rm FS}\propto t^{1/3}\propto t_{\rm obs}^{1}$ Then, the observed SSC flux roughly increases as $\nu F_\nu\propto\Gamma^3R^4\propto t_{\rm obs}^{15}$. This index is surprisingly consistent with the recently reported steep rising index of early TeV emission for GRB221009A \citep{2023Sci...380.1390L}. This could be the first evidence of the magnetic acceleration seen in GRB afterglows. 

Since the acceleration phase is at most $\sim10$ s, the keV--MeV components are probably overwhelmed by the prompt emission. Since the very high-energy components may experience $\gamma\gamma$ absorption in the prompt phase, the very early TeV afterglows might be important tracers for the acceleration phase. 
At present, only a few TeV afterglows have been confirmed \citep{2019Natur.575..455M,2019Natur.575..464A,2021Sci...372.1081H,2023Sci...380.1390L}. We expect the Cherenkov Telescope Array (CTA) will detect a lot of TeV afterglows in the next decade, a part of which may show a steep rise in the TeV band. 

\section{Summary} \label{sec:summary}

In the context of the GRB afterglow, we study the effects of the magnetization of the ejecta on the dynamics and radiation by 1D ideal relativistic MHD simulations for a wide range of the magnetization parameter as $\sigma_0=0.01$--$10$ and the initial thickness $\Delta_0/R_{\rm dec}=0.001$--$0.01$. 
We have confirmed that the magnetic effects can modify the forward and reverse shock dynamics and radiations. 
From the simulation results, we have constructed analytical formulae of the shock Lorentz factor evolution, with which we can model the afterglow light curves even for parameter ranges difficult to treat in numerical simulations.

The initial acceleration phase lasts until the reverse shock ignition time. For the magnetized thin shell case, the magnetic field efficiently accelerates the ejecta as $\Gamma\propto t^{1/3}$. Interestingly, this behaviour produces the steep SSC flux evolution $F_\nu\propto t_{\rm obs}^{15}$, which is similar to the observed early TeV light curve in GRB 221009A.

After the acceleration phase, the Lorentz factor evolution of the forward shock can be divided into three phases -- the coasting phase, the transition phase, and the BM phase. The deceleration time $t_{\rm dec}$ is an indicative parameter for the energy equipartition time, but the actual transition to the BM phase occurs later when the rarefaction wave catches up with the forward shock front. 

In the transition phase, the deceleration of the forward shock is suppressed by the effective energy injection from the shocked ejecta, resulting in $\Gamma_{\rm FS}\propto t^{1/2}$. The shock evolution slightly depends on $\sigma_0$ in the transition phase.
{The transition phase in the thick shell case can produce a flat $F_\nu\propto t_{\rm obs}^{-0.1}$ light curve lasting $10^{3.5}$ s, which is consistent with the observed characteristics of X-ray shallow decay phase.} Even with a reasonable thickness of the ejecta, a longer catch-up time of the rarefaction wave than the reverse shock crossing time can reproduce the duration of the shallow decay phase without a fine-tuned engine activity.

The ignition time of the reverse shock emission corresponds to the start of the coasting phase of the forward shock. 
Until the reverse shock crossing time $t_\Delta$, the Lorentz factor evolution of the reverse shock is almost the same as that of the forward shock.
The reverse shock crossing time is well approximated by the analytical expression Eq. (\ref{eq:RS_cross}).
The weak energy dissipation by the reverse shock for high $\sigma_0$ makes the observed flux lower. The peak time of the reverse shock light curve is regulated by the reverse shock crossing time.  
As we have demonstrated, the duration and flux of the reverse shock emission depend on the magnetization. The separation of the forward shock and reverse shock components in the lightcurve data could be crucial to determining the magnetization of the ejecta.

{Our model in this study will be tested by future observations. An increase of detection for the very rapid rising of TeV afterglow by LHAASO and CTA may provide a hint of the magnetic acceleration of the ejecta. The simultaneous detection of X-rays and GeV $\gamma$-rays in the shallow-decay phase with reverse shock components in optical bands will constrain an ejecta width and magnetization, which justifies the existence of our proposed transition phase. }

\section*{Acknowledgements}

The authors are grateful to J. Granot for useful comments on the manuscript.
The authors thankfully acknowledge the computer resources provided by the Institute for Cosmic Ray Research (ICRR), the University of Tokyo. 
This work is supported by the joint research program of ICRR, and JSPS KAKENHI Grant Numbers JP23KJ0692 (Y.K.), and JP22K03684, JP23H04899, 24H00025 (K.A.). 


\section*{Data Availability}

The data underlying this article will be shared on reasonable request to the corresponding author.



\bibliographystyle{mnras}
\bibliography{main} 




\appendix


\bsp	
\label{lastpage}
\end{document}